\documentclass[a4paper,11pt]{article}
\pdfoutput=1 

\usepackage{aas_macros} 

\usepackage{jcappub} 

\usepackage[T1]{fontenc} 

\usepackage{fontawesome} 
\usepackage{arydshln}
\usepackage{xspace}
\usepackage{todonotes} 

\usepackage{booktabs, multirow} 
\usepackage{soul}

\definecolor{deepred}{rgb}{0.8,0.2,0.2}

\definecolor{deepblue}{rgb}{0.2,0.2,0.8}

\definecolor{linkcolor}{rgb}{0.7752941176470588, 0.22078431372549023, 0.2262745098039215}
\newcommand{\nbicon}{{\color{linkcolor}\faFileCodeO}\xspace}
\newcommand{\nblink}[1]{\href{https://github.com/mkongsore/BlipFinder/tree/main/#1}{\nbicon}}
\newcommand{\githubmaster}{\href{https://github.com/mkongsore/BlipFinder}{\faGithub}\xspace}

\title{Detecting Dark Compact Objects in \textit{Gaia} DR4:
A Data Analysis Pipeline for Transient Astrometric Lensing Searches}

\author[a,1]{I-Kai Chen,\note{Co-lead author.}}
\author[a,1]{Marius Kongsore,}
\author[a,b]{and Ken Van Tilburg}

\affiliation[a]{Center for Cosmology and Particle Physics, Department of Physics,
New York University, New York, NY 10003, USA}
\affiliation[b]{Center for Computational Astrophysics, Flatiron Institute, New York, NY 10010, USA}

\emailAdd{ic2127@nyu.edu}
\emailAdd{mkongsore@nyu.edu}
\emailAdd{kenvt@nyu.edu}

\abstract{
The \textit{Gaia} satellite is cataloging the astrometric properties of an unprecedented number of stars in the Milky Way with extraordinary precision. This provides a gateway for conducting extensive surveys of transient astrometric lensing events caused by dark compact objects. In this work, we establish a data analysis pipeline capable of searching for such events in the upcoming \textit{Gaia} Data Release 4 (DR4). We use \textit{Gaia} Early Data Release 3 (EDR3) and current dark matter and astrophysical black hole population models to create mock DR4 catalogs containing stellar trajectories perturbed by lensing. Our analysis of these mock catalogs suggests that \textit{Gaia} DR4 will contain about 4 astrometric lensing events from astrophysical black holes at a $5\sigma$ significance level. Furthermore, we project that our data analysis pipeline applied to \textit{Gaia} DR4 will result in leading constraints on compact dark matter in the mass range $1$--$10^3~M_\odot$ down to a dark matter fraction of about one percent.}

\begin{document}

\maketitle
\flushbottom

\newpage

\section{Introduction}
\label{sec:intro}

A wealth of information about our universe and galaxy is contained in the spectrum of its density fluctuations and the gravitational influence they exert on other objects. All evidence for dark matter (DM) is, so far, of this kind: gravitational back-reaction on the cosmic microwave background, large-scale structure formation, cluster- and galaxy-scale velocities, and weak gravitational lensing on extra-galactic scales. From these and other indirect gravitational probes, we have learned about our cosmological history and the properties of DM and astrophysical systems on large scales.

There also exists a ``dark world'' on small scales. Most types of compact objects, such as astrophysical black holes (BHs), neutron stars, white dwarfs, brown dwarfs, and planets generically emit or reflect too little electromagnetic radiation to be detected directly, except if they are young, close, and/or accreting. This dark world may also be populated by small DM structures, such as (ultra-compact) minihalos~\cite{delos2018ultracompact,ricotti2009new,arvanitaki2020large,buschmann2020early,blinov2021dark} or more exotic objects such as primordial black holes (PBHs)~\cite{carr2020primordial}, boson stars~\cite{kaup1968klein,ruffini1969systems,schunck2003general,braaten2019colloquium,hardy2015big}, and other composite DM objects~\cite{braaten2019colloquium, frieman1988primordial, kusenko1998supersymmetric, detmold2014dark,hardy2015signatures}. These clumps and structures may be invisible to us, but their presence can occasionally be revealed indirectly through gravitational waves~\cite{raidal2017gravitational,Abbott_2018}, direct gravitational effects on visible stars~\cite{buschmann2018stellar,Andrews_2019,https://doi.org/10.48550/arxiv.2110.05549,Janssens_2022}, pulsar timing arrays~\cite{Siegel_2007,Seto_2007,Clark_2015,Schutz_2017,Baghram_2011,Kashiyama_2012,Kashiyama_2018,Dror_2019,Ramani_2020}, and gravitational lensing of light~\cite{dai2020gravitational,Alcock_1993,Alcock_1998,Alcock_2000,Blaineau_2022} and of gravitational waves~\cite{dai2018detecting,PhysRevD.106.023018,Basak_2022,10.1093/mnras/stac2944}.

Time-domain, astrometric, weak gravitational lensing of light has emerged as one of the most promising probes of compact objects in the Milky Way (MW)~\cite{van2018halometry,dominik2000astrometric,belokurov2002astrometric,erickcek2011astrometric, li2012new,vattis2021deep,mishra2022inferring,Sahu_2017_2,Kains_2017,Zurlo_2018,McGill_2022,Lu_2016}. Following the foundational works proposing the astrometric and photometric observables of transient gravitational lensing~\cite{1986ApJ...304....1P,1995A&A...294..287H,1995AJ....110.1427M,1995ApJ...453...37W}, most observational efforts have relied primarily on photometric signatures in the strong lensing regime~\cite{1991ApJ...371L..63P,1991ApJ...372L..79G,1991ApJ...374L..37M}, e.g.~managing to exclude PBHs comprising the totality of the DM abundance over a wide mass range~\cite{2019NatAs...3..524N,2019PhRvD..99h3503N,wyrzykowski2009ogle,calchi2009large,wyrzykowski2010ogle,wyrzykowski2011ogle,wyrzykowski2011ogle2, zumalacarregui2018limits}. The power of \emph{astrometric} signatures has recently received an enormous boost from simultaneous advances in catalog size, observational cadence frequency, and positional precision of astrometric surveys, most notably that of the \textit{Gaia} satellite~\cite{prusti2016gaia,brown2021gaia}, with great prospects for astrometric microlensing~\cite{2018MNRAS.476.2013R} and already many photometrically detected events~\cite{2022arXiv220606121W}, in addition to interesting candidate events from other surveys~\cite{2016MNRAS.458.3012W}. The (weak) astrometric gravitational lensing deflection signature decouples more slowly with increasing impact parameter than the (strong) photometric magnification, so it has parametric advantages in searches for rare dark objects~\cite{dominik2000astrometric}. Ref.~\cite{van2018halometry} proposed a host of observables for time-domain astrometric weak lensing by dark objects: matched filters~\cite{mondino2020first}, and correlation functions or power spectra~\cite{mishra2020power} of lensing-induced, correlated proper motion and acceleration corrections for many stars, or \emph{transient astrometric deflections} of single (or multiple) stars. 

In this paper, we present a robust data analysis pipeline to extract significant events of transient astrometric lensing on single stars, along with associated software tools and a procedure to generate faithful mock catalogs of compact objects in the MW. Our pipeline is developed with \textit{Gaia}'s fourth data release (DR4) in mind, but is applicable with minor modifications to other astrometric data sets (e.g.~HSTPROMO~\cite{van2013local} and~PHAT~\cite{dalcanton2012panchromatic}). Our robust and near-optimal data analysis pipeline is projected to detect several isolated astrophysical BHs in the MW (and perhaps other compact remnants such as neutron stars and white dwarfs), while having leading sensitivity to compact DM objects with masses between $1\,M_\odot$ and $10^3\,M_\odot$.


In related work, ref.~\cite{Verma} expanded on the sensitivity estimates of ref.~\cite{van2018halometry} by projecting the sensitivity of \textit{Gaia} time series data to PBHs using a probabilistic model, and ref.~\cite{Jab_o_ska_2022} reported the possible presence of a dark point-like lens in the observation of a single Gaia Data Release 3 (DR3) source based on a poor astrometric model fit. In this work, we faithfully produce mock data sets mimicking \textit{Gaia} DR4 time series data which include not just statistical noise, but also backgrounds from astrophysical BHs and from binary systems. Additionally, we create an analysis pipeline that can be applied to \textit{Gaia} DR4.
%

In section~\ref{sec:dynamics}, we review the basics of astrometric observations and data products in \textit{Gaia}, and how they can be affected by lensing dynamics. Section~\ref{sec:catalog} details the generation of our realistic mock catalogs, while section~\ref{sec:analysis} describes the steps in our data analysis. The results of data analyses on our mock catalogs are presented in section~\ref{sec:results}, and we conclude in section~\ref{sec:conclusions}. Supporting materials such as derivations, extra plots, and minor results can be found in appendices~\ref{sec:extended}--\ref{sec:pbh_estimate}. The data and code are available on GitHub (\githubmaster), with links (\nbicon) below each figure.

\section{Lensing dynamics}
\label{sec:dynamics}

We primarily use two models of astrometric motion in our proposed search for dark compact objects in the \textit{Gaia} DR4 data. We call the first the \textit{free model}. It describes the apparent motion of a source moving across the sky without being subject to any gravitational effects, neither local nor along the line of sight (astrometric gravitational lensing). For trajectories across small patches of the sky, this motion can be modeled as entirely inertial. The second type of model, which we call the \textit{blip model}, describes the apparent motion of a source subject to lensing due to a massive compact foreground object. By comparing the goodness-of-fit of these two models to any given source trajectory in the \textit{Gaia} catalog, we may quantitatively probe various compact DM scenarios, as well as discover singular dark compact objects in the real \textit{Gaia} data.

\subsection{Free model}
\label{sec:free_model}
We analytically model the apparent astrometric motion of an unlensed or ``free'' source across the sky, as well as the motion of point-like lenses, as a function of five parameters. The model we employ is the angular component of the ``standard model'' of stellar motion described in refs.~\cite{lindegren2012astrometric, klioner2003practical}. The angular barycentric coordinates of a free point-like celestial body  $\pmb{\theta}_{\text{free}}=(\alpha_\text{free}^*, \delta_\text{free})$ in the standard barycentric celestial reference system (BCRS, \cite{soffel2003iau}) at any given time $t$ (with respect to some fixed reference time $t_0$) are given by
\begin{equation}
\label{eq:free_source_motion}
    \pmb{\theta}_{\text{free}}(t\,|\,\pmb{\theta}_0,\pmb{\mu},D)= \pmb{\theta}_{0}+\pmb{\mu}(t-t_0)+\pmb{ \varpi}(t\,|\,\pmb{\theta}_0,D),
\end{equation}
where $\pmb{\theta}_{0}$ is the BCRS parallax subtracted position of the body at reference time $t_0$, $\pmb{\mu}=(\mu_{\alpha^*},\mu_\delta)$ is the constant angular velocity of the body in the sky, and $\pmb{\varpi}(t)$ is the parallax correction to the linear trajectory given by
\begin{equation}
\label{parallax}
    \pmb{\varpi}(t\,|\,\pmb{\theta}_0,D)=\frac{1}{D}
  \begin{bmatrix}
    \sin(\alpha) & -\cos(\alpha) & 0 \\
    \cos(\alpha)\sin(\delta) & \sin(\alpha)\sin(\delta) & -\cos(\delta)
    \end{bmatrix}
    \pmb{x}_{E,\text{cart}}(t).
\end{equation}
Here, $D$ is the line of sight distance to the object, and $\pmb{x}_{E,\text{cart}}(t)$ are the Cartesian coordinates of Earth in the heliocentric frame, which we assume to follow a purely elliptical trajectory. We note that eq.~\eqref{eq:free_source_motion} is equivalent to the 5-parameter astrometric model \textit{Gaia} use to model each source trajectory that they measure. Hence the set of parameters $(\pmb{\theta}_0,\pmb{\mu},D)$ are the same as reported by \textit{Gaia} in all data releases thus far, except \textit{Gaia} uses parallax $\varpi$ as a parameter instead of distance $D$. The two are equivalent since $D=(1\;\text{arcsec}/\varpi) \;\text{pc}$.\footnote{With the well-known caveat that the inferred parallax may be negative for distant or poorly measured stars, leading to an unphysical (negative) distance. This failure mode is eliminated by imposing priors.} An example of a free trajectory can be seen in figure~\ref{fig:example_trajectories}.


We additionally model source trajectories undergoing constant angular acceleration. We achieve this by adding two extra parameters to the free model
\begin{equation}
\label{eq:accel_source_motion}
    \pmb{\theta}_{\text{accel}}(t\,|\,\pmb{\theta}_0,\pmb{\mu},D)= \pmb{\theta}_{\text{free}}(t\,|\,\pmb{\theta}_0,\pmb{\mu},D)+\frac{1}{2}\pmb{\gamma}(t-t_0)^2
\end{equation}
where $\pmb{\gamma}=(\gamma_{\alpha^*},\gamma_\delta)$ is the constant angular acceleration of the celestial body in the sky. We use the acceleration model to discriminate between long period binaries and blips. See section~\ref{sec:binaries} for more details.

\subsection{Blip model}

We model the trajectory of a celestial body subject to detectable transient astrometric lensing caused by a point-like lens as a function of 11 parameters. We call this the ``blip'' model of celestial motion~\cite{van2018halometry}. In addition to the 5 free motion parameters of eq.~\eqref{eq:free_source_motion}, there are 6 additional parameters: the position of the lens $\pmb{\theta}_{l,0}$ at reference time $t_0$, the proper motion of the lens $\pmb{\mu}_l$, the distance to the lens $D_l$, and the mass of the lens $m_l$. In its most basic form, the blip model may be written as
\begin{equation}
    \pmb{\theta}_{\text{blip}}(t) = \pmb{\theta}_{\text{free}}(t) + \Delta\pmb{\theta}(t).
\label{eq:general_blip_trajectory}
\end{equation}
We calculate the lensing deflection term $\Delta\pmb{\theta}(t)$ assuming a point-like lens, and we employ the thin-lens approximation, in which we assume the lensing deflection takes place over a region that is very small compared to the line-of-sight distances involved. The point-like lens assumption allows us to construct a model that is valid in both the weak and strong lensing regimes. These approximations are valid as long as the Newtonian potential of the lens is small and the relative velocities of the observer, lens, and source are small compared to the speed of light, which is the case for all sources in the \textit{Gaia} catalog.

\begin{figure}
\begin{center}
\tikzset{every picture/.style={line width=0.75pt}} 
\hspace*{-1.0in}
\begin{tikzpicture}[x=0.75pt,y=0.75pt,yscale=-1,xscale=1]

\draw [color={rgb, 255:red, 245; green, 166; blue, 35 }  ,draw opacity=1 ]   (23.55,62.45) -- (374.68,102.04) ;
\draw [color={rgb, 255:red, 245; green, 166; blue, 35 }  ,draw opacity=1 ]   (23.55,62.45) -- (373.83,22.67) ;
\draw [line width=0.75]    (27,142) -- (512,142) ;
\draw [shift={(515,142)}, rotate = 180] [fill={rgb, 255:red, 0; green, 0; blue, 0 }  ][line width=0.08]  [draw opacity=0] (5.36,-2.57) -- (0,0) -- (5.36,2.57) -- (3.56,0) -- cycle    ;
\draw [shift={(24,142)}, rotate = 0] [fill={rgb, 255:red, 0; green, 0; blue, 0 }  ][line width=0.08]  [draw opacity=0] (5.36,-2.57) -- (0,0) -- (5.36,2.57) -- (3.56,0) -- cycle    ;
\draw    (27,111.99) -- (371.87,111.36) ;
\draw [shift={(374.87,111.36)}, rotate = 179.9] [fill={rgb, 255:red, 0; green, 0; blue, 0 }  ][line width=0.08]  [draw opacity=0] (5.36,-2.57) -- (0,0) -- (5.36,2.57) -- (3.56,0) -- cycle    ;
\draw [shift={(24,112)}, rotate = 359.9] [fill={rgb, 255:red, 0; green, 0; blue, 0 }  ][line width=0.08]  [draw opacity=0] (5.36,-2.57) -- (0,0) -- (5.36,2.57) -- (3.56,0) -- cycle    ;
\draw  [dash pattern={on 4.5pt off 4.5pt}]  (374.87,62.36) ;
\draw  [color={rgb, 255:red, 245; green, 166; blue, 35 }  ,draw opacity=1 ][fill={rgb, 255:red, 255; green, 255; blue, 255 }  ,fill opacity=1 ] (510.2,32.06) .. controls (510.2,30.05) and (511.83,28.42) .. (513.84,28.42) .. controls (515.85,28.42) and (517.48,30.05) .. (517.48,32.06) .. controls (517.48,34.07) and (515.85,35.7) .. (513.84,35.7) .. controls (511.83,35.7) and (510.2,34.07) .. (510.2,32.06) -- cycle ;
\draw    (6,62) -- (26.67,52.33) ;
\draw    (6,62) -- (26.67,72.33) ;
\draw [color={rgb, 255:red, 245; green, 166; blue, 35 }  ,draw opacity=1 ] [dash pattern={on 4.5pt off 4.5pt}]  (373.83,22.67) -- (511.58,7.61) ;
\draw [shift={(513.91,7.36)}, rotate = 353.76] [color={rgb, 255:red, 245; green, 166; blue, 35 }  ,draw opacity=1 ][line width=0.75]      (0, 0) circle [x radius= 3.35, y radius= 3.35]   ;
\draw [color={rgb, 255:red, 245; green, 166; blue, 35 }  ,draw opacity=1 ]   (373.83,22.67) -- (513.84,32.06) ;
\draw [color={rgb, 255:red, 245; green, 166; blue, 35 }  ,draw opacity=1 ]   (513.84,32.06) -- (374.68,102.04) ;
\draw [shift={(513.84,32.06)}, rotate = 153.3] [color={rgb, 255:red, 245; green, 166; blue, 35 }  ,draw opacity=1 ][fill={rgb, 255:red, 245; green, 166; blue, 35 }  ,fill opacity=1 ][line width=0.75]      (0, 0) circle [x radius= 3.35, y radius= 3.35]   ;
\draw [color={rgb, 255:red, 245; green, 166; blue, 35 }  ,draw opacity=1 ] [dash pattern={on 4.5pt off 4.5pt}]  (374.68,102.04) -- (512.28,118.15) ;
\draw [shift={(514.62,118.42)}, rotate = 6.68] [color={rgb, 255:red, 245; green, 166; blue, 35 }  ,draw opacity=1 ][line width=0.75]      (0, 0) circle [x radius= 3.35, y radius= 3.35]   ;
\draw  [draw opacity=0] (358.29,24.88) .. controls (359.82,30.31) and (360.95,35.87) .. (361.66,41.53) -- (192.95,55.4) -- cycle ; \draw    (359.08,27.82) .. controls (360.22,32.32) and (361.08,36.89) .. (361.66,41.53) ;  \draw [shift={(358.29,24.88)}, rotate = 77.31] [fill={rgb, 255:red, 0; green, 0; blue, 0 }  ][line width=0.08]  [draw opacity=0] (7.14,-3.43) -- (0,0) -- (7.14,3.43) -- cycle    ;
\draw  [draw opacity=0] (316.12,43.95) .. controls (318.13,56.39) and (319.16,69.07) .. (319.14,81.94) .. controls (319.14,86.31) and (319.01,90.65) .. (318.77,94.98) -- (-78.45,81.45) -- cycle ; \draw    (316.12,43.95) .. controls (318.13,56.39) and (319.16,69.07) .. (319.14,81.94) .. controls (319.14,85.3) and (319.06,88.65) .. (318.92,91.99) ; \draw [shift={(318.77,94.98)}, rotate = 271.77] [fill={rgb, 255:red, 0; green, 0; blue, 0 }  ][line width=0.08]  [draw opacity=0] (7.14,-3.43) -- (0,0) -- (7.14,3.43) -- cycle    ; 
\draw  [draw opacity=0] (349.37,42.45) .. controls (350.48,49.02) and (351.25,55.66) .. (351.68,62.38) -- (37.55,74.45) -- cycle ; \draw    (349.86,45.51) .. controls (350.71,51.08) and (351.32,56.71) .. (351.68,62.38) ;  \draw [shift={(349.37,42.45)}, rotate = 82.18] [fill={rgb, 255:red, 0; green, 0; blue, 0 }  ][line width=0.08]  [draw opacity=0] (7.14,-3.43) -- (0,0) -- (7.14,3.43) -- cycle    ;
\draw [color={rgb, 255:red, 245; green, 166; blue, 35 }  ,draw opacity=1 ][fill={rgb, 255:red, 255; green, 255; blue, 255 }  ,fill opacity=1 ] [dash pattern={on 4.5pt off 4.5pt}]  (23.55,62.45) -- (511.49,32.2) ;
\draw [shift={(513.84,32.06)}, rotate = 356.45] [color={rgb, 255:red, 245; green, 166; blue, 35 }  ,draw opacity=1 ][line width=0.75]      (0, 0) circle [x radius= 3.35, y radius= 3.35]   ;
\draw  [draw opacity=0] (23.24,54.26) .. controls (23.85,56.66) and (24.15,59.17) .. (24.09,61.76) .. controls (24.02,64.69) and (23.5,67.5) .. (22.59,70.14) -- (-4.52,61.11) -- cycle ; \draw   (23.24,54.26) .. controls (23.85,56.66) and (24.15,59.17) .. (24.09,61.76) .. controls (24.02,64.69) and (23.5,67.5) .. (22.59,70.14) ;  
\draw  [dash pattern={on 4.5pt off 4.5pt}]  (23.55,62.45) -- (374.87,62.36) ;
\draw [shift={(374.87,62.36)}, rotate = 359.98] [color={rgb, 255:red, 0; green, 0; blue, 0 }  ][fill={rgb, 255:red, 0; green, 0; blue, 0 }  ][line width=0.75]      (0, 0) circle [x radius= 3.35, y radius= 3.35]   ;
\draw [shift={(23.55,62.45)}, rotate = 359.98] [color={rgb, 255:red, 0; green, 0; blue, 0 }  ][fill={rgb, 255:red, 0; green, 0; blue, 0 }  ][line width=0.75]      (0, 0) circle [x radius= 3.35, y radius= 3.35]   ;

\draw (175,94.4) node [anchor=north west][inner sep=0.75pt]    {$D_{l}$};
\draw (174,124.4) node [anchor=north west][inner sep=0.75pt]    {$D_{s}$};
\draw (66,64) node [anchor=north west][inner sep=0.75pt]   [align=left] {};
\draw (371,67.4) node [anchor=north west][inner sep=0.75pt]    {$l$};
\draw (521,27.4) node [anchor=north west][inner sep=0.75pt]    {$s$};
\draw (523,2.4) node [anchor=north west][inner sep=0.75pt]    {$s_{+}$};
\draw (522,113.4) node [anchor=north west][inner sep=0.75pt]    {$s_{-}$};
\draw (335,45.4) node [anchor=north west][inner sep=0.75pt]    {$\beta $};
\draw (363,23.4) node [anchor=north west][inner sep=0.75pt]    {$\Delta \theta _{+}$};
\draw (322.74,78.35) node [anchor=north west][inner sep=0.75pt]    {$\Delta \theta _{-}$};
\draw (7,70.4) node [anchor=north west][inner sep=0.75pt]    {$o$};

\end{tikzpicture}

\end{center}

\caption{Astrometric lensing geometry. A point-like lens $l$ at a line-of-sight distance $D_l$ from an observer $o$ creates two displaced images $s_+$ and $s_-$ of a background source $s$ at an angular impact parameter $\mathbf{\beta}$ and line-of-sight distance $D_s$. The displaced images are separated from the true source location $s$ by angles $\Delta\theta_+$ and $\Delta\theta_-$ as specified by eq.~\eqref{eq:two_soln_lensing_defl}. We average the location of $s_+$ and $s_-$ weighted by their relative magnification to obtain a single lensed source location.}
\label{fig:lensing_diagram}
\end{figure}
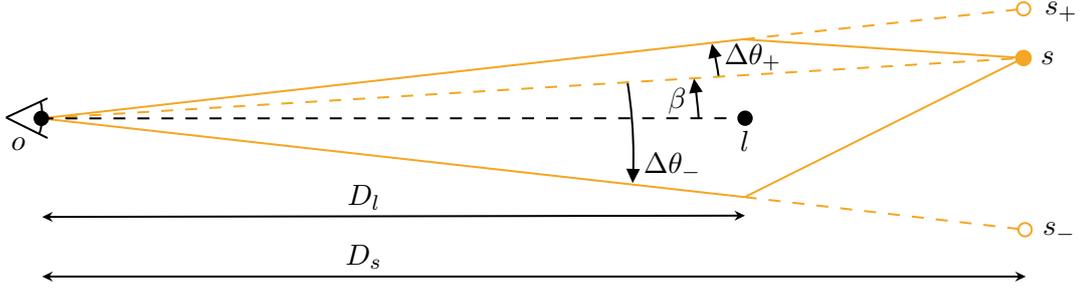

\begin{figure}
    \centering
    \includegraphics[width=\textwidth,height=\textheight,keepaspectratio]{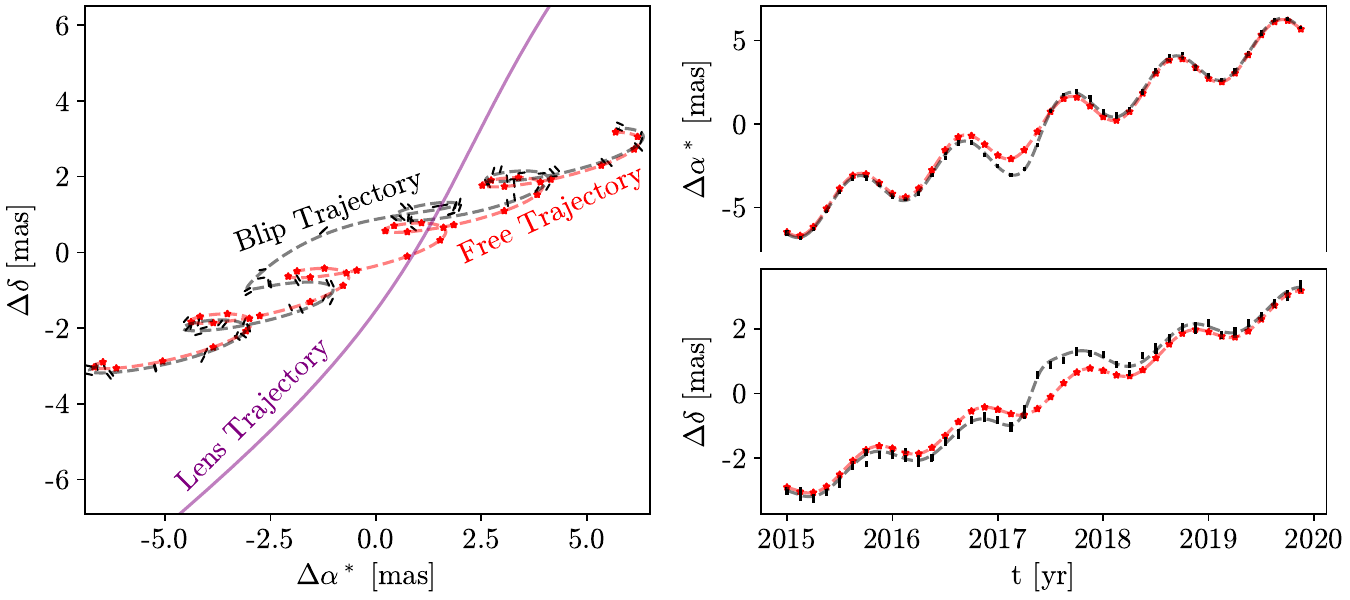}
    \caption{A mock blip event at $(\alpha,\delta)=(62.35^\circ,34.5^\circ)$. The source is at a distance of $1200\;\text{pc}$ from the Solar System and is being deflected by an $8\; M_\odot$ lens at a distance of $800\;\text{pc}$, with minimal angular impact parameter of $|\pmb{\beta}|_{\text{min}}=0.092\;\text{mas}$. \textit{Left:} In red, the free trajectory of the source, with each star marking the location of the source at each \textit{Gaia} epoch, and the dashed line marking the continuous trajectory of the source. In solid purple, the lens trajectory. In dashed black, the deflected trajectory of the source, with mock data points and corresponding error bars rotated to point along the \textit{Gaia} AL scan angle. \textit{Upper Right:} The free and deflected right ascension coordinates of the source as a function of time. \textit{Lower Right:} The free and deflected declination coordinates of the source as a function of time. \nblink{PaperPlots/source_dynamics_plot.ipynb}}
    \label{fig:example_trajectories}
\end{figure}

The Einstein radius $\theta_{E}$ of a massive point-like object is given by
\begin{equation}
\label{eq:einstein_radius}
    \theta_{E} = 2\sqrt{\frac{G m_l}{c^2}\bigg(\frac{D_s - D_l}{D_l  D_s}\bigg)}\approx 2.85\; \text{mas}\;\left(\frac{m_l}{M_\odot}\right)^{\frac12}\left(\frac{1\;\text{kpc}}{D_l}\right)^{\frac12}\left(\frac{D_s-D_l}{D_s}\right)^{\frac12},
\end{equation}
where $D_s$ and $D_l$ are the distance to the source and lens, respectively, $m_l$ is the mass of the lens, $G$ is the gravitational constant, and $c$ is the speed of light~\cite{1992grle.book.....S}. 
Using the Einstein radius, we then calculate the deflection of the two images created by the lens as
\begin{equation}
    \Delta\pmb{\theta}_{\pm}(t) = \frac{1}{2}\bigg(\pm\sqrt{|\pmb{\beta}(t)|^2+4\theta_E^2}-|\pmb{\beta}(t)|\bigg)\hat{\pmb{\beta}},
\label{eq:two_soln_lensing_defl}
\end{equation}
with relative (signed) magnification
\begin{equation}
    \mu_\pm(t)=\Bigg| 1-\bigg(\frac{\theta_E}{|\Delta\pmb{\theta}_{\pm}(t)+\pmb{\beta}(t)|}\bigg)^4 \Bigg|^{-1}=\frac{u^2(t)+2}{2u(t)\sqrt{u^2(t)+4}}\pm \frac{1}{2},
\label{eq:two_soln_lensing_mag}
\end{equation}
where $\pmb{\beta}(t)$ is the angular impact parameter pointing from the lens to the source, and the dimensionless impact parameter $u(t)\equiv|\pmb{\beta}(t)/\theta_E|$. The absolute value on the left hand side accounts for the fact that the inversion of the second image can be ignored since it is point-like. Given \textit{Gaia}'s point spread function (PSF) width of about 2~pixels or $100\,\mathrm{mas}$~\cite{prusti2016gaia}, the two lensed source images are rarely resolved individually (especially if the lensing occurs inside the MW), meaning \textit{Gaia} will usually only resolve the light centroid of the two images. Via eqs.~\eqref{eq:two_soln_lensing_defl} and~\eqref{eq:two_soln_lensing_mag}, the light centroid deflection due to lensing is given by
\begin{equation}
    \Delta\pmb{\theta}(t)=\theta_E\frac{u(t)}{u^2(t)+2}\hat{\pmb{\beta}},
\label{eq:lensing_defl}
\end{equation}
which we insert into eq.~\eqref{eq:general_blip_trajectory} to obtain a complete expression for \emph{lensed} trajectories in the \textit{Gaia} catalog. We provide a schematic of the lensing geometry and notation in figure~\ref{fig:lensing_diagram}, and we show a realistic blip trajectory in figure~\ref{fig:example_trajectories}.

\begin{figure}
    \centering
    \includegraphics[width=0.6\textwidth,height=0.6\textheight,keepaspectratio]{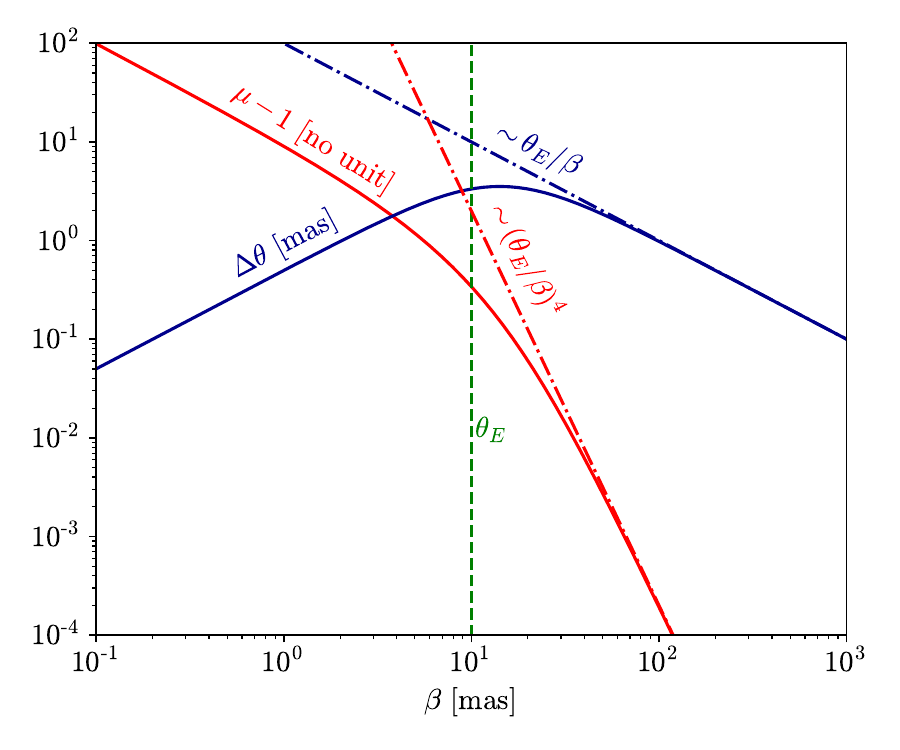}
    \caption{Photometric and astrometric observables for a single lensing event, caused by a lens with a 10.0 mas Einstein radius passing by a luminous star. In solid red, magnification observable $\mu-1$ as a function of impact parameter, with the asymptotic behavior ($\mu-1\sim (\theta_E/\beta)^4$ for $\beta\gg\theta_E$) shown in dashed red. In solid blue, astrometric deflection observable $\Delta\theta$ as a function of impact parameter, with the asymptotic behavior ($\Delta\theta\sim \theta_E/\beta$ for $\beta\gg\theta_E$) shown in dashed blue. The lens Einstein radius is shown in green. \nblink{PaperPlots/astrometry_vs_photometry_plot.ipynb}}
\label{fig:observable_behavior}
\end{figure}

By contrast, adding the two image magnifications in \eqref{eq:two_soln_lensing_mag}, we obtain a total magnification of
\begin{equation}
    \mu(t) = \frac{u^2(t)+2}{u\sqrt{u^2+4}},
\label{eq:total_mag}
\end{equation}
which leads to an effective change in source magnitude of $\Delta$Mag $=-2.5\log_{10}(1/\mu)$. In figure~\ref{fig:observable_behavior}, we show the the astrometric deflection given by~\eqref{eq:lensing_defl} and the photometric magnification given by~\eqref{eq:total_mag} as a function of impact parameter, as well as the asymptotic behavior of each observable, for a gravitational lens with Einstein radius $\theta_E=10.0$~mas.

\section{Mock catalog}
\label{sec:catalog}

\indent In this section, we describe our method for creating mock catalogs that closely resemble the data products from the upcoming \textit{Gaia} DR4. First, we discuss how to extrapolate the 5-parameter astrometric solution reported by \textit{Gaia} EDR3 into the time-series data expected in \textit{Gaia} DR4. Then, we describe the models we adopted for generating astrophysical BHs and compact DM. The mock catalog provides a way to understand the statistical background for event selection, detectable lensing events, and the projected compact DM constraints which are shown in section~\ref{sec:results}.

\subsection{\textit{Gaia} EDR3 extrapolation}
We take all the sources in \textit{Gaia} EDR3 that have a 5-parameter astrometric solution and generate time-series data in the proposed format of \textit{Gaia} DR4. By using astrometric parameters and stellar magnitudes directly from EDR3, we automatically capture extinction, crowding, and instrumental effects, which normally must be treated carefully in catalogs based on the injection of fully artificial stars. Some of the stars in EDR3 have negative parallaxes and large parallax uncertainty. To circumvent this issue, we take the median of the inferred distance posterior of each star with geometric and photometric priors prescribed in ref.~\cite{Bailer_Jones_2021}. With the unlensed mock catalog, we can test the false positive rate for lensing events and determine the distribution of our test statistics under the null hypothesis. We also inject lenses using astrophysically realistic priors on their phase space distribution to construct a lensed catalog.

The epoch astrometry due to be released in DR4 will not provide timestamped two-dimensional BCRS coordinates due to the scanning law of \textit{Gaia}~\cite{prusti2016gaia}. Instead, each epoch measurement will be reported as a one-dimensional displacement $\theta(t)$ with respect to a scan angle $\phi(t)$ in the so-called ``Along Scan Direction'' (AL) in the \textit{Gaia} documentation.
We convert the coordinates given by our model to this data format using the relation
\begin{equation}
\label{eq:scan_angle_formula}
    \theta(t) =
    [\alpha(t)-\alpha_0]\sin\phi(t)+[\delta(t)-\delta_0]\cos\phi(t),
\end{equation}
\noindent where ($\alpha_0,\delta_0$) are the BCRS coordinates of the source at a reference time $t_0$ provided by \textit{Gaia}. Only the brightest stars will have a location offset in the perpendicular ``Across Scan Direction'' (AC). For simplicity, we will only use the AL location for all the stars in our mock catalog. The timestamp and scan angle for each epoch will be the same for all stars in the catalog. The data points are evenly spread over 40 timestamps between the start and end of the observations covered by \textit{Gaia} DR4. (The \textit{Gaia} nominal mission time is from Jul 2014 to Jul 2019 \cite{prusti2016gaia} but we use Jan 2015 to Dec 2019 for simplicity.) An additional 40 points about two hours apart from the first set of 40 points (with the same set of scan angles) are added to the time series to mimic the scanning law described in ref.~\cite{prusti2016gaia}, for a combined total of 80 data points.

We note that our pipeline is also capable of using \textit{Gaia}'s Observation Forecast Tool (GOST) to obtain more accurate scan angles and observation timestamps for each source.\footnote{The \textit{Gaia}'s Observation Forecast Tool (\url{https://gaia.esac.esa.int/gost/}) provides a forecast of \textit{Gaia} observations and scan angles.} In appendix~\ref{sec:gost}, we further discuss GOST and show limits on dark compact objects obtained using a mock catalog generated with GOST, analogous to the limits shown in section~\ref{sec:results}. We also discuss how GOST affects \textit{Gaia}'s ability to discover BHs. We emphasize that these limits and \textit{Gaia}'s discovery potential are only marginally different to the ones obtained in the simplified data scenario where all sources are observed exactly 80 times. Therefore, all subsequent sections assume this simplified scenario.


\subsection{Lens populations}
\label{sec:populations}
We inject isolated, electromagnetically quiet BHs and compact DM objects into the mock catalog. The priors for generating these two different populations are specified in the following.

\subsubsection{Astrophysical BHs}
\label{sec:bh_pop}
MW stellar evolution simulations suggest that there should be of order $10^8$ BHs in the MW \cite{Olejak_2020}, yet we have only observed a handful through the emission of electromagnetic waves from accretion and photometric microlensing. \textit{Gaia} DR4 will provide an opportunity to discover isolated, non-accreting BHs via transient astrometric lensing.

Since astrophysical BHs are remnants of stellar evolution, we assume that their distribution in the sky closely resembles the MW stellar distribution. The stellar population in the MW is commonly decomposed into the Galactic bulge, thin disk, thick disk, and the Galactic halo. The thin disk is of primary relevance for our purposes, due to its high stellar number density and its proximity to Earth. We model the Galactic thin disk with the exponential function
\begin{equation}
\label{eq:thin_disk}
    n_*(R,z) = n_0\exp\left(-\frac{|z|}{z_d} - \frac{R}{R_d}\right),
\end{equation}
where $n_0$ is the central stellar number density, and $z_d$ and $R_d$ are the scale height and scale radius of the thin disk, respectively. Ref.~\cite{McMillan_2016} reports $z_d = 300$ pc and $R_d = 2.6$ kpc.

\begin{figure}
    \centering
    \includegraphics[width = \textwidth]{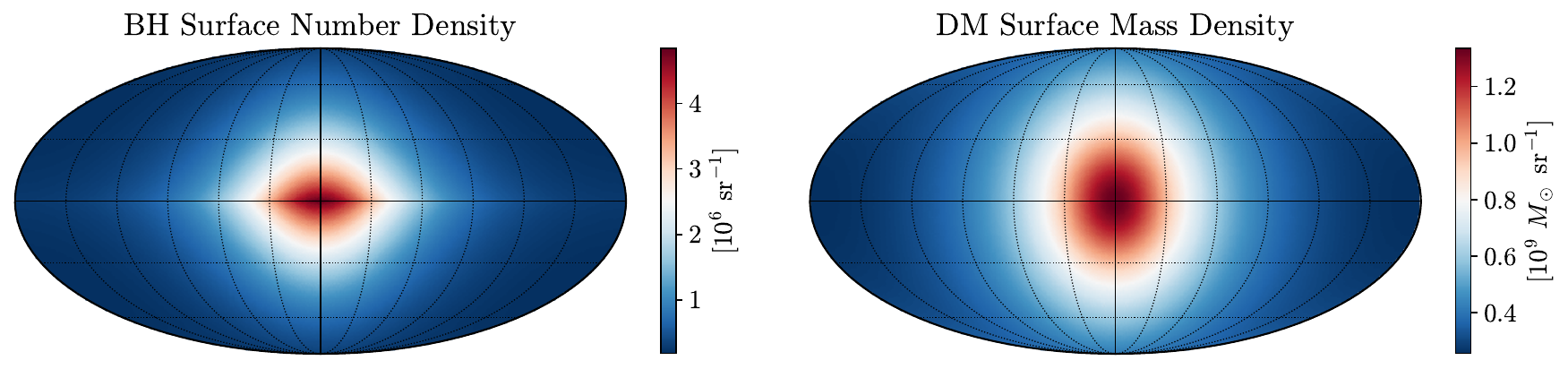}
    \caption{\textit{Left:} BH surface number density obtained by integrating the BH volume number density inside a sphere of radius 5~kpc centered on the solar system, taken to be 8~kpc away from the Galactic Center.
    We assume there are a total of $10^8$ BHs present in the entire Galactic thin disk \cite{Olejak_2020}, and $10^7$ within 5~kpc of the Sun.
    \textit{Right:} DM surface mass density obtained by integrating the BH volume mass  density inside a sphere of radius 5~kpc centered on the solar system. We assume the MW DM density follows an NFW profile of scale radius 18~kpc with a value $10^{-2}~M_\odot$/pc$^3$ at the Sun's location. \nblink{PaperPlots/BH_DM_density.ipynb}}
    \label{fig:BH_abundance}
\end{figure}

Simply using stellar distributions to model the MW BH distribution does not account for BH natal kicks. These kicks --- caused by the dynamics of supernova explosions --- offset the BH velocity distribution from that of MW stars. These kicks explain the observed distribution of low mass X-ray binaries far away from the Galactic disk \cite{Repetto_2012,Janka_2013}, because the BH velocity gain due to kicks will increase the scale height $z_d$ of the BH distribution relative to that of the stellar distribution, effectively ``puffing up'' the disk. We estimate in appendix~\ref{sec:bh_pm_pdf} that the scale height will increase by a factor of about $10$ due to this effect, so for astrophysical BHs, we use $z_d = 3$ kpc and $R_d = 2.6$ kpc. The surface number density of BHs across the sky is shown in figure~\ref{fig:BH_abundance}. The probability density function (PDF) for BH distances $D_l$ at a given celestial location in galactic coordinates $(l,b)$ is then
\begin{equation}
    P_\text{BH}(D_l| l, b) \propto D_l^2~n_\text{BH}\Big(R(D_l, l, b), z(D_l, b)\Big).
    \label{eq:distance_prior}
\end{equation}
The combined PDF of BH proper motion and distance is
\begin{equation}
    P_\text{BH}(\pmb{\mu}_l, D_l| l, b) =P_\text{BH}(\pmb{\mu}_l|D_l, l, b) P_\text{BH}(D_l| l, b),
    \label{eq:pm_d_prior}
\end{equation}
which we normalize such that $\int P_\text{BH}(\pmb{\mu}_l, D_l| l, b)\mathrm{d^2}\mu_l\mathrm{d}D_l = 1$. For a detailed derivation of the conditional PDF $P_\text{BH}(\pmb{\mu}_l|D_l, l, b)$, see appendix~\ref{sec:bh_pm_pdf}.

\begin{figure}
    \centering
    \includegraphics[width = 0.6\textwidth]{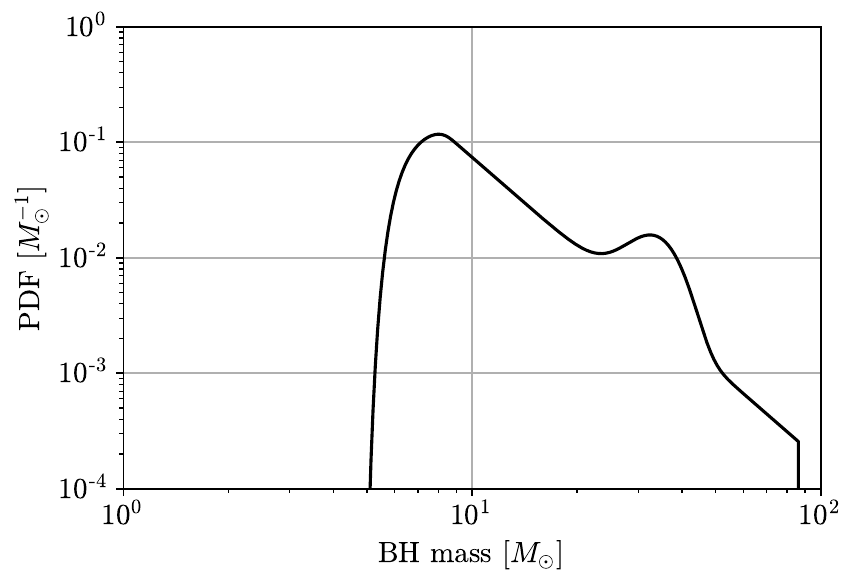}
    \caption{The BH mass function adapted from the LIGO-Virgo \textsc{Power Law $+$ Peak} model \cite{Abbott:2020gyp}. The lower bound is at $4.59 ~M_\odot$, the upper bound is at $86.22 ~ M_\odot$. The peak is a Gaussian centered at $33.07 ~M_\odot$ with standard deviation $5.69 ~M_\odot$. \nblink{PaperPlots/BH_mass.ipynb}}
    \label{fig:BH_mass}
\end{figure}

We adopt the BH mass distribution reported by LIGO-Virgo~\cite{Abbott:2020gyp} obtained from a combination of 47 binary BH merger observations. We thus assume --- for now --- that the BH mass distribution is similar for single BHs and for binary BHs.\footnote{One of the derived end products of our data analyses on \textit{Gaia} DR4 and other data sets will be to pin down the mass function for \emph{isolated} astrophysical BHs.} We also assume that the BH mass is independent of the position and the proper motion of the BH so that the two PDFs $P_\text{BH}(\pmb{\mu}_l, D_l| l, b), P_\text{BH}(M_\text{BH})$ are separable.  The model we use is the \textsc{Power Law $+$ Peak} model reported by LIGO-Virgo, wherein the BH mass distribution follows a power law with a soft cutoff at the lower end and a hard cutoff at the upper end. A peak is added, motivated by a potential pile up of BHs just before the pair-instability gap of supernovae~\cite{2018ApJ...856..173T}. The resulting BH mass function is shown in figure~\ref{fig:BH_mass}.

\subsubsection{Compact DM objects}
\label{sec:PBH}

Compact DM objects may comprise part or all of the DM abundance and thus produce transient astrometric lensing signals in \textit{Gaia} DR4. A non-detection would set constraints on the fraction of DM composed of such compact objects (e.g.~PBHs) as a function of their mass. Here, we only consider point-like sources, specifically lens objects with scale radii smaller than their Einstein radius
\begin{equation}
    r_s<D_l\theta_E\approx1.38\times10^{-5}~\text{pc}~ (2.85~\text{AU}) \left(\frac{m_l}{M_\odot}\right)^\frac{1}{2}\left(\frac{D_l}{1 \text{ kpc}}\right)^\frac{1}{2}\left(\frac{D_s-D_l}{D_s}\right)^{\frac12}.
\end{equation}
(In appendix~\ref{sec:extended}, we discuss the limitations on detecting lensing events from lenses with extended density profiles.) We assume that the DM distribution in the MW follows a Navarro–Frenk–White (NFW) profile \cite{NFW} with a fiducial scale radius $R_s = 18~$kpc and a local DM density $\rho_\odot = 10^{-2}M_\odot/\text{pc}^3$ \cite{McMillan_2016}
\begin{equation}
    \rho_\text{NFW}(r) = \frac{\rho_0}{\frac{r}{R_s}\left(1 + \frac{r}{R_s}\right)^2},
\end{equation}
and that the DM has a Gaussian velocity distribution
\begin{equation}
    P(\mathbf{v_\text{DM}}) = \frac{1}{(2\pi\sigma_\text{DM}^2)^{3/2}}\exp\left(-\frac{\mathbf{v_\text{DM}}^2}{2\sigma_\text{DM}^2}\right),
\end{equation}
where $\sigma_\text{DM} = 166~$km/s. The surface mass density of DM across the sky is shown in figure~\ref{fig:BH_abundance}. With these parameters, there is roughly $7.0\times10^{10} \, M_\odot$ of DM mass within a 13~kpc radius around the solar system, corresponding to the 99th percentile of the stellar distances in our mock \textit{Gaia} DR4 catalog (based on EDR3).

\subsection{Noise}
\label{sec:noise}
\begin{figure}

    \centering
    \includegraphics[width=\textwidth,height=\textheight,keepaspectratio]{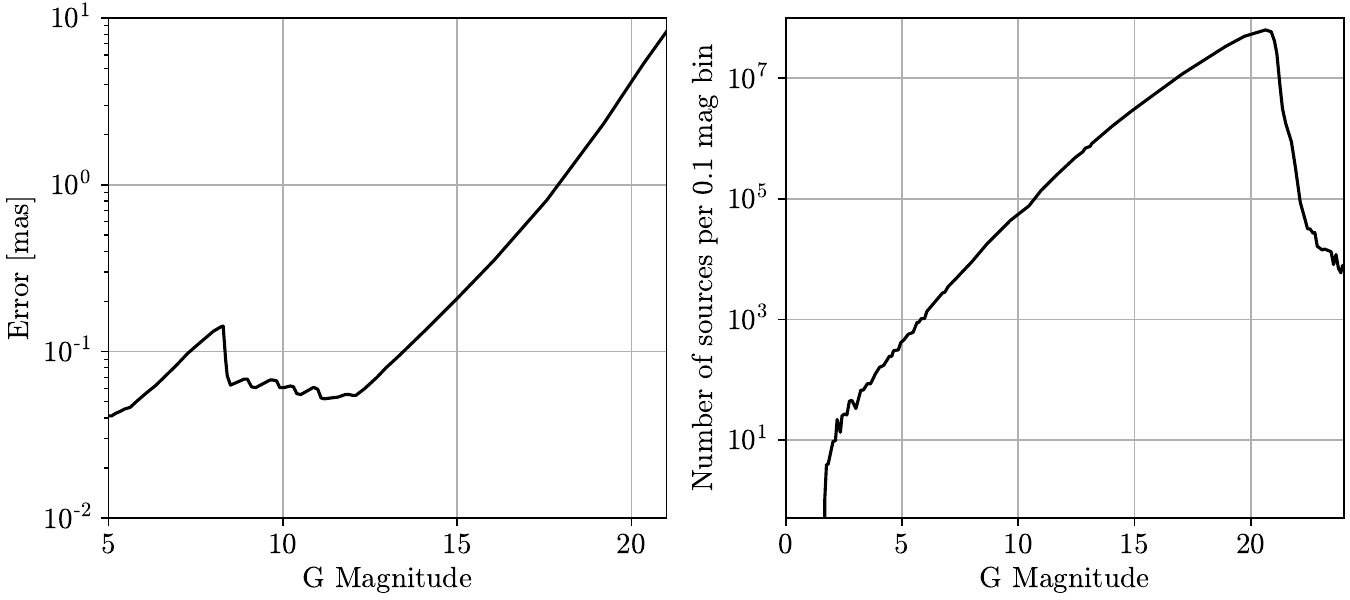}
    \caption{\textit{Left:} The \textit{Gaia} EDR3 error function~\cite{Lindegren_2021}, showing the median per transit astrometric error of a given source as a function of G magnitude. \textit{Right:} The \textit{Gaia} EDR3 G magnitude distribution~\cite{Lindegren_2021}. Since we construct our mock catalogs based on EDR3, the astrometric errors are distributed exactly according to these two distributions. \nblink{PaperPlots/error_plot.ipynb}}
    \label{fig:error}
\end{figure}

We perturb each astrometric positional data point generated via the free and blip models by subjecting the mock source trajectories to Gaussian noise. Since we base our mock catalogs on \textit{Gaia} EDR3, we draw directly from the EDR3 error distribution. In practice, this is done by using the error function described in ref.~\cite{Lindegren_2021} to convert each EDR3 source's reported photometric mean G magnitude into a Gaussian standard deviation quantifying the instrumental astrometric precision in the AL scan direction for a single transit. We then randomly shuffle each positional data point in every source trajectory by drawing from a normal distribution centered at each true source position and with standard deviation corresponding to the per transit error. The EDR3 error function and the EDR3 G magnitude distribution are shown in figure~\ref{fig:error}. Note that using the EDR3 error function is conservative, since errors are projected to decrease in future data releases across all G magnitudes~\cite{esa_cosmos}. For simplicity, the error function we use here is only a function of the stellar magnitude. In reality, the error function is position dependent as shown in ref.~\cite{Lindegren_2021}. In crowded regions, such as inside the Galactic Bulge, uncertainties may be larger than for stars of similar magnitude located outside the bulge. However, since we directly apply the median \textit{Gaia} EDR3 error function to our analysis, these error anisotropies should not significantly affect the final results reported in section~\ref{sec:results}.

\section{Data analysis}
\label{sec:analysis}

In this section, we describe our construction of a data analysis pipeline to detect true blip events and set constraints on dark compact object populations in both the true \textit{Gaia} DR4 catalog and the mock catalogs described in section~\ref{sec:catalog}. The pipeline systematically goes through an entire catalog and optimizes a set of test statistics for each source in order to discern the probability that any given source trajectory is a true blip event. By making cuts in the significance level of different test statistics, we can thus discriminate between blip and free stellar trajectories, and thus discover and flag true blip events effectively. We can also obtain limits on the compact object DM fraction in the MW using the Yellin method \cite{Yellin_2002,Yellin_2008} applied on the distribution of these test statistics.

\subsection{Blip test statistics}
As pointed out in section~\ref{sec:noise}, we assume the astrometric \textit{Gaia} DR4 data to be subject to pure Gaussian noise, with the positional error of each source corresponding to its G magnitude. Hence, we use a Gaussian likelihood function to quantify the agreement between the astrometric data and our choice of model (either free or blip). Given a dataset $\pmb{\theta}_\text{obs} = \lbrace \theta_{n,\text{obs}} \rbrace$ where the subscript $n$ labels each data point in the source trajectory, as well as either 5 parameters $\pmb{y} = \pmb{y}_\text{free}$ (free model) or 11 parameters $\pmb{y} = \pmb{y}_\text{blip}$ (blip model), we may write the corresponding likelihood function as
\begin{equation}
    \mathcal{L}(\pmb{\theta}_\text{obs}|\pmb{y})= \prod_n \frac{1}{\sqrt{2\pi}{{\sigma}_n}} \exp\bigg(-\frac{(\theta_{n,\text{obs}}-\theta_{n,\text{model}})^2}{2 \sigma_n^2} \bigg)
    \label{eq:lhood}
\end{equation}
where $\pmb{\theta}_\text{model}(\pmb{y}) = \lbrace \theta_{n,\text{model}} \rbrace$ is the prediction given the model parameters $\pmb{y}$, and $\sigma_n$ is the error associated with the data point $n$. We then define our blip \textit{test statistic} ($\text{TS}$) to be
\begin{equation}
\label{eq:ts}
    \text{TS}(\pmb{\theta}_\text{obs})\equiv -2\bigg[\max_{\pmb{y}_{\text{free}}}\log\mathcal{L}(\pmb{\theta}_\text{obs} |\pmb{y}_{\text{free}})-\max_{\pmb{y}_{\text{blip}}}{\log \mathcal{L}(\pmb{\theta}_\text{obs} |\pmb{y}_{\text{blip}})}\bigg],
\end{equation}
namely, the test statistic for any given source trajectory is defined as the maximized log likelihood ratio between the free and blip model fits to the source trajectory data. We note that the negative log likelihood ratio is equivalent to the difference in $\chi^2$ goodness of fit values between the two models. It should also be noted that under the assumption of trivial covariance between model parameters, the distribution of maximized test statistics follows a true $\chi^2$ distribution in the asymptotic limit~\cite{Cowan_2011}.

While eq.~\eqref{eq:ts} provides a way to evaluate the quality of fit of our model to the data, the expression does not contain any prior information on the lens population being probed. To constrain our search, we therefore construct a second test statistic based on the posterior of a lensing event, rather than the likelihood. We define this \textit{constrained} test statistic ($\text{TS}^*$) as
\begin{equation}
\label{eq:ts_constrained}
    \text{TS}^*(\pmb{\theta}_\text{obs})\equiv -2\bigg[{\max_{\pmb{y}_{\text{free}}}}\, \log\mathcal{L}(\pmb{\theta}_\text{obs}  |\pmb{y}_{\text{free}})-{\max_{\pmb{y}_{\text{blip}}}}^*\,{\log\mathcal{L}(\pmb{\theta}_\text{obs} |\pmb{y}_{\text{blip}})}\bigg],
\end{equation}
where $\max^*$ indicates that rather than maximizing the blip likelihood directly, we are instead maximizing the log of the posterior probability associated with each source trajectory
\begin{equation}
\label{eq:posterior}
    P_{\text{post}} = \log\big[\mathcal{L}(\pmb{\theta}_\text{obs} |\pmb{y}_{\text{blip}})P(\pmb{y}_\text{lens})\big],
\end{equation}
where $P(\pmb{y}_{\text{lens}})$ is the prior probability density of the lens parameters, with the exact form of the prior depending on the lens population being probed, as described in section~\ref{sec:populations}. Note that the quantity inside the square brackets has nontrivial units, but these can be neglected since they amount to a constant offset in the test statistic and hence do not matter if eq.~\eqref{eq:posterior} is used as a loss function only. A further constraint implied by $\max^*$ is the requirement
\begin{equation}
    \text{blippiness}(\pmb{y}_\text{blip})\equiv \frac{t_\text{obs}\mu_\text{rel}}{\beta_\text{min}}>1,
\label{eq:blippiness}
\end{equation}
where $\mu_\text{rel}$ is the relative (linear) proper motion magnitude between the source and the lens, $t_\text{obs}$ is the total observation time, and $\beta_\text{min}$ is the minimum angular impact parameter between the lens and the source. We have coined the above quantity the \textit{blippiness} of an event, as it is simply the ratio between the relative angular distance traversed by the source and lens over the full observation time $\tau$, and the minimum angular impact parameter.

There are two reasons for imposing these extra constraints when maximizing the log likelihood ratio. First, maximizing the posterior rather than the likelihood means that we penalize choices of model parameters that are unphysical. Similarly, were we not to impose the blippiness constraint, we would be probing parts of parameter space which cannot produce a significant blip, simply because events that have a large minimal impact parameter are either too long or the lensing deflection is too weak to produce a signal. Second, imposing these constraints guides our choice of minimizer to a physical part of the blip parameter space, which reduces the amount of computational power needed to compute test statistics for all $2\times10^9$ events in the \textit{Gaia} catalog.

We emphasize that constraining the maximization in eq.~\eqref{eq:ts_constrained} only reduces the value of the test statistic compared to what would be obtained by calculating eq.~\eqref{eq:ts}, meaning the full test statistic distribution gets shifted to smaller (or even negative) values. However, for true blip events, the reduction in significance is minimal due to the distribution of true blip parameters coinciding with the prior probability distribution in eq.~\eqref{eq:posterior}.

Finally, we note that our analysis pipeline is also capable of incorporating the prior on stellar distances reported in ref.~\cite{Bailer_Jones_2021}. This prior helps overcome \textit{Gaia}'s difficulty in determining stellar parallaxes for faint sources or sources located in the galactic bulge, where effects from blending and crowding can be significant. We do not include this prior in the analysis reported here; however, we tested how it affects the results reported in section~\ref{sec:results} and found no significant difference. However, for the real DR4 data set, where bad parallax measurements have a more significant impact on the analysis, the Bailer-Jones prior will be beneficial. It is therefore enabled by default in the analysis software.

\subsection{Constraining compact DM objects}
\label{sec:constrain_method}
We employ the optimum interval method developed by Yellin \cite{Yellin_2002, Yellin_2008} to determine (projected) limits on the DM fraction $f_l$ in compact DM objects. The Yellin method is suited to hypothesis testing of a known signal model in the presence of an \emph{unknown} background distribution, in a fixed region of interest. For a one-dimensional distribution of \emph{events}, it entails computing the integral of the \emph{signal} distribution of all intervals of $n$ events and assesses whether the largest interval significantly exceeds the expectation for the signal model, in which case the signal hypothesis is rejected.

In our analysis, the \emph{events} are the constrained test statistics for all of the stars. We can compute the distribution of test statistics under the signal (lensing) hypothesis numerically by drawing compact DM objects from the distributions specified in section~\ref{sec:PBH}. For computational efficiency, we only consider stars in the distribution whenever a lens is present within a threshold impact parameter which causes a maximum deflection of at least $\Delta \theta = 5~\mu$as.

The background distribution is obtained by fitting the unlensed catalog; the \emph{background events} are the large upwards statistical fluctuations in the constrained test statistics. Furthermore, we can consider a mock catalog contaminated with lensing by astrophysical BHs, and by binary systems with an undetected companion as astrophysical backgrounds.

The recipe of implementing the optimal interval method in this work is the following:
\begin{enumerate}
    \item Generate a test statistic distribution only for stars that have a nearby lens. We call it the ``signal distribution'' $S(\text{TS})$.
    \item Given the test statistics of the experiment, compute the maximum of expected number of events between all pairs of events $e_i, e_{i+(n+1)}$, which is the integral of $S(\text{TS})$ between $e_i, e_{i+(n+1)}$. We call this the \emph{maximum interval} $x_n$.
    \item Generate many instances of Monte Carlo realizations of the signal events and perform step 2 on all of the realizations.
    \item Compute the probability that the $x_n$ in the experiment is larger than the Monte Carlo realizations. We call this probability $C_n$. Compute the maximum of $C_n$,  $C_\text{Max} = \max_n\{C_n\}$.
    \item Repeat step 4 comparing each Monte Carlo realization with all other realizations and calculate their $C_\text{Max}$. If the $C_\text{Max}$ from the experiment is larger than $90\%$ of the $C_\text{Max}$ of the Monte Carlo realization, then we say the signal model is rejected at $90\%$ confidence level.
\end{enumerate}
The DM fraction $f_l$ is simply a scaling factor in the signal distribution $S(\text{TS})$. Following the steps outlined above, we find the limiting $f_{l,*}$ such that the DM fraction $f_l\geq f_{l.*}$ is excluded at $90\%$ confidence level. For a more detailed discussion on the Yellin method, see refs.~\cite{Yellin_2002, Yellin_2008}.

\subsection{Analysis pipeline}
\label{sec:pipeline}

\textit{Gaia} DR4 will contain time series data for about 2 billion sources. Scouring this vast catalog for blip events is a considerable computational challenge and requires a structured approach. We construct a modular analysis pipeline wherein key statistical assumptions, such as the lens priors, can be swapped to search for blips from different lens populations. Ancillary data from e.g. photometric surveys can also be incorporated via these priors.

\begin{figure}[!h]
    \centering
    \tikzset{every picture/.style={line width=0.75pt}} 
    \resizebox{1.0\textwidth}{!}{%
    \begin{tikzpicture}[x=0.75pt,y=0.75pt,yscale=-1,xscale=1]
    \hspace*{-3em}
    \path (0,489); 

    \draw    (497,400.33) -- (546.67,432) ;
    \draw [shift={(526.05,418.85)}, rotate = 212.52] [fill={rgb, 255:red, 0;     green, 0; blue, 0 }  ][line width=0.08]  [draw opacity=0] (8.93,-4.29) --     (0,0) -- (8.93,4.29) -- cycle    ;
    \draw    (440.67,400.67) -- (387.67,432) ;
    \draw [shift={(409.86,418.88)}, rotate = 329.41] [fill={rgb, 255:red, 0;     green, 0; blue, 0 }  ][line width=0.08]  [draw opacity=0] (8.93,-4.29) --     (0,0) -- (8.93,4.29) -- cycle    ;
    \draw    (468.67,340.67) -- (468.67,360.67) ;
    \draw [shift={(468.67,355.67)}, rotate = 270] [fill={rgb, 255:red, 0;     green, 0; blue, 0 }  ][line width=0.08]  [draw opacity=0] (8.93,-4.29) --     (0,0) -- (8.93,4.29) -- cycle    ;
    \draw    (418.74,270.94) -- (468.28,301.86) ;
    \draw [shift={(447.75,289.05)}, rotate = 211.97] [fill={rgb, 255:red, 0;     green, 0; blue, 0 }  ][line width=0.08]  [draw opacity=0] (8.93,-4.29) --     (0,0) -- (8.93,4.29) -- cycle    ;
    \draw    (418.74,210.94) -- (418.67,230.67) ;
    \draw [shift={(418.68,225.8)}, rotate = 270.21] [fill={rgb, 255:red, 0;     green, 0; blue, 0 }  ][line width=0.08]  [draw opacity=0] (8.93,-4.29) --     (0,0) -- (8.93,4.29) -- cycle    ;
    \draw    (369.74,140.94) -- (420.67,171.54) ;
    \draw [shift={(399.49,158.82)}, rotate = 211] [fill={rgb, 255:red, 0;     green, 0; blue, 0 }  ][line width=0.08]  [draw opacity=0] (8.93,-4.29) --     (0,0) -- (8.93,4.29) -- cycle    ;
    \draw    (98.74,70.94) -- (149.33,101.33) ;
    \draw [shift={(128.32,88.71)}, rotate = 211] [fill={rgb, 255:red, 0; green,     0; blue, 0 }  ][line width=0.08]  [draw opacity=0] (8.93,-4.29) -- (0,0) --     (8.93,4.29) -- cycle    ;
    \draw  [dash pattern={on 4.5pt off 4.5pt}]  (99,141) .. controls     (-90.38,273.48) and (263.62,241.48) .. (99,351) ;
    \draw [shift={(87.06,253.27)}, rotate = 207.25] [fill={rgb, 255:red, 0;     green, 0; blue, 0 }  ][line width=0.08]  [draw opacity=0] (8.93,-4.29) --     (0,0) -- (8.93,4.29) -- cycle    ;
    \draw  [dash pattern={on 4.5pt off 4.5pt}]  (319.08,140.55) .. controls     (69.16,259.85) and (424.62,256.48) .. (228.25,351.22) ;
    \draw [shift={(245.85,248.8)}, rotate = 221.8] [fill={rgb, 255:red, 0;     green, 0; blue, 0 }  ][line width=0.08]  [draw opacity=0] (8.93,-4.29) --     (0,0) -- (8.93,4.29) -- cycle    ;
    \draw  [fill={rgb, 255:red, 255; green, 255; blue, 255 }  ,fill opacity=1 ]     (19,39) .. controls (19,34.58) and (22.58,31) .. (27,31) -- (171.17,31) ..     controls (175.59,31) and (179.17,34.58) .. (179.17,39) -- (179.17,63) ..     controls (179.17,67.42) and (175.59,71) .. (171.17,71) -- (27,71) ..     controls (22.58,71) and (19,67.42) .. (19,63) -- cycle ;
    \draw  [fill={rgb, 255:red, 255; green, 255; blue, 255 }  ,fill opacity=1 ]     (70,109) .. controls (70,104.58) and (73.58,101) .. (78,101) --     (222.17,101) .. controls (226.59,101) and (230.17,104.58) .. (230.17,109)     -- (230.17,133) .. controls (230.17,137.42) and (226.59,141) ..     (222.17,141) -- (78,141) .. controls (73.58,141) and (70,137.42) ..     (70,133) -- cycle ;
    \draw  [fill={rgb, 255:red, 255; green, 255; blue, 255 }  ,fill opacity=1 ]     (290,109) .. controls (290,104.58) and (293.58,101) .. (298,101) --     (442.17,101) .. controls (446.59,101) and (450.17,104.58) .. (450.17,109)     -- (450.17,133) .. controls (450.17,137.42) and (446.59,141) ..     (442.17,141) -- (298,141) .. controls (293.58,141) and (290,137.42) ..     (290,133) -- cycle ;
    \draw  [fill={rgb, 255:red, 255; green, 255; blue, 255 }  ,fill opacity=1 ]     (340,179) .. controls (340,174.58) and (343.58,171) .. (348,171) --     (492.17,171) .. controls (496.59,171) and (500.17,174.58) .. (500.17,179)     -- (500.17,203) .. controls (500.17,207.42) and (496.59,211) ..     (492.17,211) -- (348,211) .. controls (343.58,211) and (340,207.42) ..     (340,203) -- cycle ;
    \draw  [fill={rgb, 255:red, 255; green, 255; blue, 255 }  ,fill opacity=1 ]     (340,239) .. controls (340,234.58) and (343.58,231) .. (348,231) --     (492.17,231) .. controls (496.59,231) and (500.17,234.58) .. (500.17,239)     -- (500.17,263) .. controls (500.17,267.42) and (496.59,271) ..     (492.17,271) -- (348,271) .. controls (343.58,271) and (340,267.42) ..     (340,263) -- cycle ;
    \draw  [fill={rgb, 255:red, 255; green, 255; blue, 255 }  ,fill opacity=1 ]     (390,309) .. controls (390,304.58) and (393.58,301) .. (398,301) --     (542.17,301) .. controls (546.59,301) and (550.17,304.58) .. (550.17,309)     -- (550.17,333) .. controls (550.17,337.42) and (546.59,341) ..     (542.17,341) -- (398,341) .. controls (393.58,341) and (390,337.42) ..     (390,333) -- cycle ;
    \draw  [fill={rgb, 255:red, 255; green, 255; blue, 255 }  ,fill opacity=1 ]     (390,369) .. controls (390,364.58) and (393.58,361) .. (398,361) --     (542.17,361) .. controls (546.59,361) and (550.17,364.58) .. (550.17,369)     -- (550.17,393) .. controls (550.17,397.42) and (546.59,401) ..     (542.17,401) -- (398,401) .. controls (393.58,401) and (390,397.42) ..     (390,393) -- cycle ;
    \draw  [fill={rgb, 255:red, 255; green, 255; blue, 255 }  ,fill opacity=1 ]     (480,440) .. controls (480,435.58) and (483.58,432) .. (488,432) --     (640.67,432) .. controls (645.08,432) and (648.67,435.58) .. (648.67,440)     -- (648.67,464) .. controls (648.67,468.42) and (645.08,472) ..     (640.67,472) -- (488,472) .. controls (483.58,472) and (480,468.42) ..     (480,464) -- cycle ;
    \draw  [fill={rgb, 255:red, 255; green, 255; blue, 255 }  ,fill opacity=1 ]     (288.67,440) .. controls (288.67,435.58) and (292.25,432) .. (296.67,432)     -- (451,432) .. controls (455.42,432) and (459,435.58) .. (459,440) --     (459,464) .. controls (459,468.42) and (455.42,472) .. (451,472) --     (296.67,472) .. controls (292.25,472) and (288.67,468.42) .. (288.67,464)     -- cycle ;
    \draw    (229.74,120.94) -- (290.08,120.54) ;
    \draw [shift={(264.91,120.71)}, rotate = 179.62] [fill={rgb, 255:red, 0;     green, 0; blue, 0 }  ][line width=0.08]  [draw opacity=0] (8.93,-4.29) --     (0,0) -- (8.93,4.29) -- cycle    ;
    \draw    (39,351) -- (289.44,350.71) ;

    \draw (117,112) node [anchor=north west][inner sep=0.75pt]  [font=\Large]     [align=left] {Free Fit};
    \draw (316,112) node [anchor=north west][inner sep=0.75pt]  [font=\Large]     [align=left] {Free Sampling};
    \draw (434,312) node [anchor=north west][inner sep=0.75pt]  [font=\Large]     [align=left] {Blip Fit};
    \draw (414,373) node [anchor=north west][inner sep=0.75pt]  [font=\Large]     [align=left] {Blip Sampling};
    \draw (376,183) node [anchor=north west][inner sep=0.75pt]  [font=\Large]     [align=left] {Accel. Fit};
    \draw (355,241) node [anchor=north west][inner sep=0.75pt]  [font=\Large]     [align=left] {Accel. Sampling};
    \draw (317,442) node [anchor=north west][inner sep=0.75pt]  [font=\Large]     [align=left] {Yellin Bounds};
    \draw (497,443) node [anchor=north west][inner sep=0.75pt]  [font=\Large]     [align=left] {Blip Candidates};
    \draw (52,354) node [anchor=north west][inner sep=0.75pt]  [font=\large]     [align=left] {Excluded from further analysis};
    \draw (75,225) node [anchor=north west][inner sep=0.75pt]      [font=\footnotesize,rotate=-28.23]  {$-2\log\hat{\mathcal{L}}_{\rm free} < \chi^2_{5\sigma }$};
    \draw (245,219) node [anchor=north west][inner sep=0.75pt]      [font=\footnotesize,rotate=-39.74]  {$-2\log\hat{\mathcal{L}}_{\rm free}^{\prime} < \chi _{5\sigma     }^{2}$};
    \draw (233.86,100.77) node [anchor=north west][inner sep=0.75pt]      [font=\footnotesize,rotate=20]  {$-2\log\hat{\mathcal{L}}_{\rm free} \geq \chi _{5\sigma     }^{2}$};
    \draw (418.16,155.83) node [anchor=north west][inner sep=0.75pt]      [font=\footnotesize,rotate=20]  {$-2\log\hat{\mathcal{L}}_{\rm free}^{\prime} \geq \chi _{5\sigma     }^{2}$};
    \draw (41,41) node [anchor=north west][inner sep=0.75pt]  [font=\Large]     [align=left] {\textit{Gaia} Catalog};

    \end{tikzpicture}
    }%

    \caption{A flowchart representation of the analysis pipeline. The pipeline reads in astrometric data from the input catalog. Then, the free model is fitted to every source in the catalog, generating a set of optimized log likelihoods $\{-2\log\hat{\mathcal{L}}_{\rm free}\}$. Any events that are below a $5\sigma$ threshold in the free model $\chi^2$ distribution is excluded. To ensure that all free fits have converged to global minima, we then rerun the same optimization procedure, except we use an nested sampling based optimizer, yielding a new set of log likelihoods $\{-2\log\hat{\mathcal{L}}'_{\rm free}\}$. We then reimpose the $5\sigma$ cutoff on the new computed free log likelihoods. Any event that passes these cuts is then tested against the acceleration model, also using the nested sampling optimizer. Finally, we test the remaining events against the blip model by computing $\text{TS}^*$. All events that pass the initial cut, as well as a $3\sigma$ cut in acceleration, and which satisfy $\text{TS}^*>100$ are flagged as blip events. The $\text{TS}^*$ distribution for events passing the initial $5\sigma$ free cut is also returned, which the pipeline uses to impose constraints on lens populations using the Yellin method.}
\label{fig:pipeline}
\end{figure}
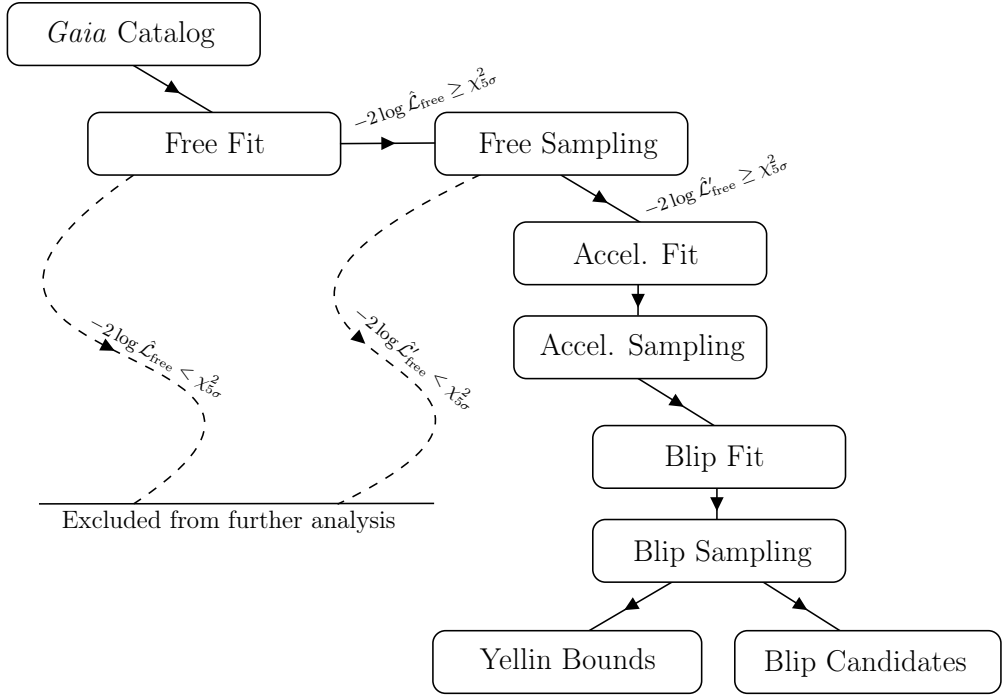

A diagrammatic representation of the analysis pipeline's flow is shown in figure~\ref{fig:pipeline}. We first fit the free model from section~\ref{sec:free_model} to every source in the catalog. To do this, we employ \texttt{SciPy}'s \texttt{minimize} function \cite{scipy} to maximize the logarithm of eq.~\eqref{eq:lhood}. This yields an optimized log likelihood $-2\log\hat{\mathcal{L}}_{\rm free}$ value for each source trajectory, where the ``hat'' indicates that the likelihood has been maximized with respect to the source trajectory. After performing the initial fit, we impose our first cut. Any \emph{significant} blip event should have a small optimal likelihood under the free trajectory hypothesis; we discard any events with an optimized negative log likelihood of $-2\log\hat{\mathcal{L}}_{\rm free}<\chi^2_{5\sigma}$, where $\chi^2_{5\sigma}$ is the $5\sigma$ significance threshold of the $-2\log\hat{\mathcal{L}}_{\rm free}$ distribution computed via Monte Carlo (MC). This distribution asymptotically matches a $\chi^2$ distribution with $m-5$ degrees of freedom, where $m$ is the number of data points in a given observation (see section~\ref{sec:unperturbed_catalog}), where $\chi^2_{5\sigma}=152$ for 75 degrees of freedom (all trajectories in the mock catalog consist of 80 data points) computed by matching the $\chi^2$ distribution to the Gaussian $5\sigma$ p-value of $5.7\times10^{-7}$.  For any events that pass this cut, we rerun the free model fit, but this time using a nested sampling procedure using the Bayesian inference tool \texttt{PyMultinest}~\cite{Feroz_2008,Feroz_2009,Feroz_2019,buchner}. This ensures that the \emph{global} maximum of each free model log likelihood is found. Should any of the remaining sources fall under the $5\sigma$ threshold after this second fit, they also get discarded. This cut yields the most significant reduction in computational resources needed to search for blips, since it reduces the number of sources of interest by 6 orders of magnitude.

We then fit sources that pass the first two cuts against the acceleration model. This extra fit is primarily implemented to account for binaries (see section~\ref{sec:binaries} for more details). Like with the free fit, we minimize $-2\log\mathcal{L}_{\rm accel}$ using first \texttt{SciPy} and then \texttt{PyMultinest}, yielding a set of optimized log likelihoods $-2\log\hat{\mathcal{L}}_{\rm accel}$.

\begin{figure}
    \centering    \includegraphics[width=\textwidth,height=\textheight,keepaspectratio]{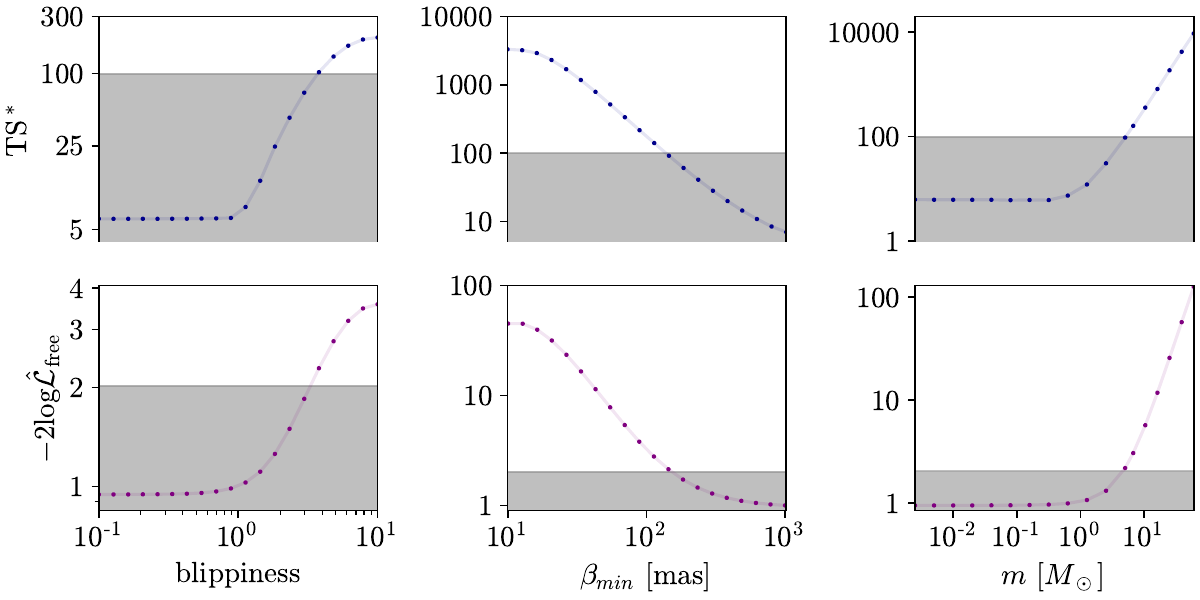}
    \caption{Single variable variation plots. \textit{Left:} The constrained test statistic and optimized free log likelihood computed as a function of event blippiness for a mock blip event with source $(\alpha_0,\delta_0) = (0^{\circ},0^{\circ})$, $(\mu_{\alpha^*},\mu_\delta)=(30,30)$~mas/yr, $d=1000~\mathrm{pc}$, and lens parameters $(\Delta\alpha_0,\Delta\delta_0) = (70.71,-70.71)$~mas, $d=500~\mathrm{pc}$, $m=7~M_\odot$, $\beta_{\rm min}= 100$~mas, and where the blippiness is varied by varying the proper motion of the lens. \textit{Middle:} Constrained test statistic and optimized free log likelihood computed as a function of minimal impact parameter, with the same source parameters as for the left hand case, but with a lens with $\text{blippiness}=6.36$, $d=500~\mathrm{pc}$, $m=7~M_\odot$, and the remaining parameters varied. \textit{Right:} Constrained test statistic and optimized free log likelihood for a blip event as a function of mass. The source parameters are again the same as in the middle and left hand case, but the lens parameters are $(\Delta\alpha_0,\Delta\delta_0) = (70.71,-70.71)$~mas, $(\mu_{\alpha^*},\mu_\delta)=(-60,-60)$~mas/yr, $d=500~\mathrm{pc}$, $\beta_{\rm min}= 100$~mas, and $\text{blippiness}=6.36$. The gray areas indicate regions in which an event meets our cut criteria and gets excluded from the list of blips found by the analysis pipeline. Note that all three of these true blip events cross the $5\sigma$ free log likelihood threshold and the $\text{TS}^*>100$ requirement at roughly the same value in the varied blip parameter. Also note that at very low blippiness or $\beta_{\rm min}$, the blip events enters the strongly lensed regime, which is the cause of the flatness of both the test statistic and likelihood in this range. \nblink{PaperPlots/single_variable_plot.ipynb}}
    \label{fig:single_variable}
\end{figure}

Finally, we compute the constrained test statistic $\text{TS}^*$ for each remaining source using again first \texttt{SciPy} and then \texttt{PyMultinest} to ensure convergence to global maxima. Any event that passes the $5\sigma$ free model cut, is above $3\sigma$ significance under the assumption of the acceleration model, and has a test statistic $\text{TS}^*>100$, is flagged as a blip event. Furthermore, for these events (and any other event that passed the initial $5\sigma$ free fit cut), the pipeline outputs a list of test statistics ($-2\log\hat{\mathcal{L}}_{\rm free}$, $-2\log\hat{\mathcal{L}}_{\rm accel}$, and $\text{TS}^*$), each model's best fit parameters and corresponding uncertainties, and nested sampling generated parameter space covariance data. See figure~\ref{fig:single_variable} for an example of the pipeline's sensitivity to changes in various lens parameters. The pipeline is finally also able to run a Yellin test on the computed $\text{TS}^*$ distribution and can generate 90\% confidence limits on compact DM parameter space.

\section{Mock results}
\label{sec:results}

We run the data analysis pipeline of section~\ref{sec:pipeline} on the mock catalogs described in section~\ref{sec:catalog} to test its ability to discover true blip events in quasi-realistic data, and to make projections for the discovery potential and expected constraints in \textit{Gaia} DR4.
 We first apply the pipeline on a mock catalog unperturbed by lensing to quantify the distribution of test statistics generated by the analysis procedure, as well as to ensure that the pipeline is robust against random noise, misfitting errors, and other artifacts. We then run it blindly on the astrophysical BH catalog described in section~\ref{sec:bh_catalog} in order to probe its ability to detect this astrophysical signal that is guaranteed to be present in the data. Next, we run the pipeline on a series of mock binary events with dark companions to ensure that the pipeline will not flag binaries with dark companions as blips. Finally, we use the pipeline on the compact DM catalog described in section~\ref{sec:PBH} to generate mock Yellin 90\% limits on the compact DM fraction in the MW.

\subsection{The unperturbed catalog}
\label{sec:unperturbed_catalog}

\begin{figure}
    \centering
    \includegraphics[width=0.7\textwidth,height=\textheight,keepaspectratio]{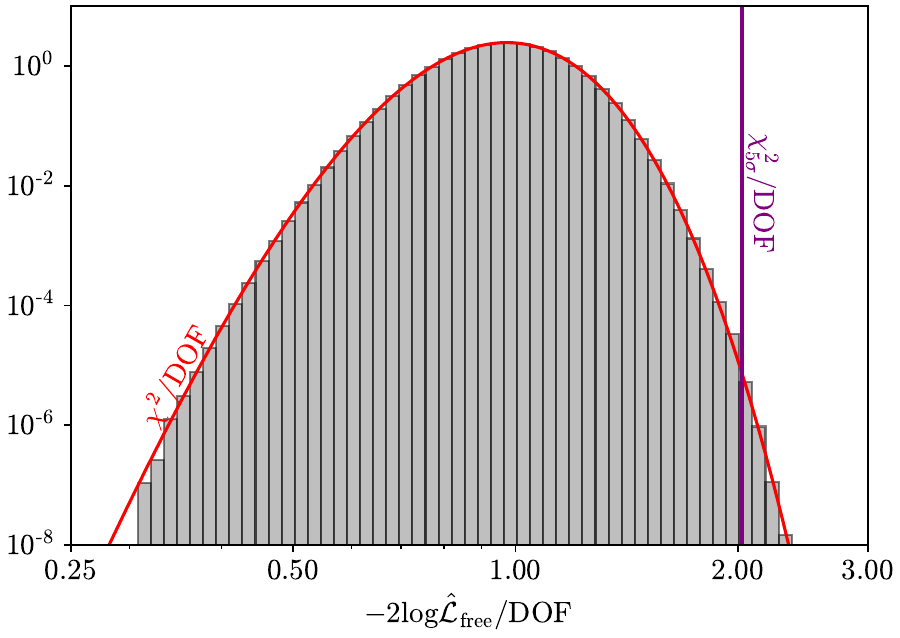}
    \caption{Normalized histogram (gray) showing the $-2\log\hat{\mathcal{L}_\text{free}}$/DOF distribution for all 1,447,353,154 sources in the unperturbed mock catalog. In red, the analytic $\chi^2$/DOF probability density function. In purple, the (upper) $5\sigma$ threshold of the analytic $\chi^2$/DOF distribution. The $\chi^2$ distribution obtained by the analysis pipeline matches the theoretical distribution nearly perfectly, which is expected since this particular catalog only contains sources with free trajectories subject to Gaussian noise. The strong agreement demonstrates the robustness of the pipeline. \nblink{PaperPlots/null_catalog_plot.ipynb}
    }
    \label{fig:x0_chisq}
\end{figure}

We first analyze the mock catalog consisting of 1,447,353,154 \textit{Gaia} sources propagating freely across the sky, without undergoing any sort of lensing deflection. Figure~\ref{fig:x0_chisq} shows the $-2\log\hat{\mathcal{L}}_\text{free}$ distribution for these events. The log likelihood distribution closely follows an analytic $\chi^{2}$ distribution, in line with expectations for Gaussian noise injection only, and highlighting  that the free model's 5 parameters have minimal covariance.

Some events in this catalog pass the initial $5\sigma$ cut in the free log likelihood distribution. This is expected from statistical noise and the sheer number of events in the catalog. Upon computing $\text{TS}^*$, however, we see that \emph{none} of the events in this catalog pass the $\text{TS}^*>100$ requirement for an event to be flagged as a blip. In fact, all of the events satisfy $\text{TS}^*<60$, meaning none of the events are even remotely close to being considered as a highly significant blip event. The stringent cuts in log likelihoods and in $\text{TS}^*$ effectively preclude statistical fluctuations from being classified as blips, at least under our assumption of high-quality data with Gaussian noise.

\subsection{The BH catalog}
\label{sec:bh_catalog}
\begin{figure}
    \centering    \includegraphics[width=\textwidth,height=\textheight,keepaspectratio]{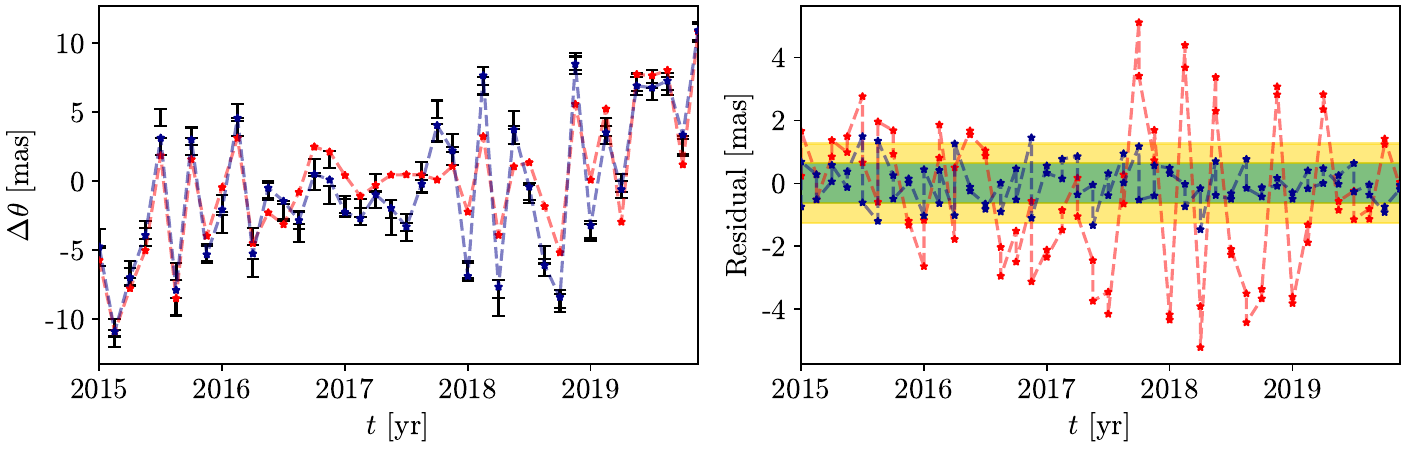}
    \caption{The best free and blip model fits to mock catalog source 5727504125199235456. \textit{Left:} The AL scan angle displacement data in black, with the best fit free model in red, and the best fit blip model in purple. Note that the dashed lines do not show the continuous trajectory of the model and data, but rather simply connect the data points since their order can otherwise be hard to gauge. \textit{Right:} The residuals from the fits, with the 1$\sigma$ and 2$\sigma$ bands being shown in green and yellow, respectively. \nblink{PaperPlots/significant_plot.ipynb}}
\label{fig:sig1}
\end{figure}

\begin{figure}
    \centering    \includegraphics[width=\textwidth,height=\textheight,keepaspectratio]{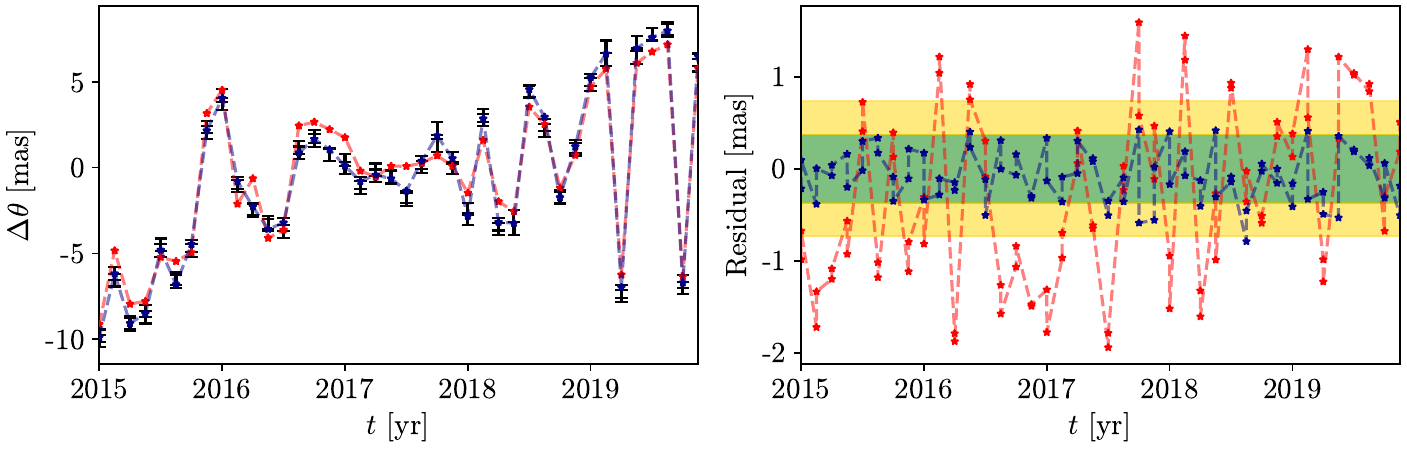}
    \caption{The best free and blip model fits to mock catalog source 6262458554071571712. \textit{Left:} The AL scan angle displacement data in black, with the best fit free model in red, and the best fit blip model in purple. \textit{Right:} The residuals from the fits, with the 1$\sigma$ and 2$\sigma$ bands being shown in green and yellow, respectively. Note that despite the large uncertainties in the parameters of this source and the other 4, the blip model still provides an excellent fit. This is due to parameter degeneracy. \nblink{PaperPlots/significant_plot.ipynb}}
\label{fig:sig2}
\end{figure}

\begin{figure}
    \label{fig:triangle}
    \centering\includegraphics[width=\textwidth,height=\textheight,keepaspectratio]{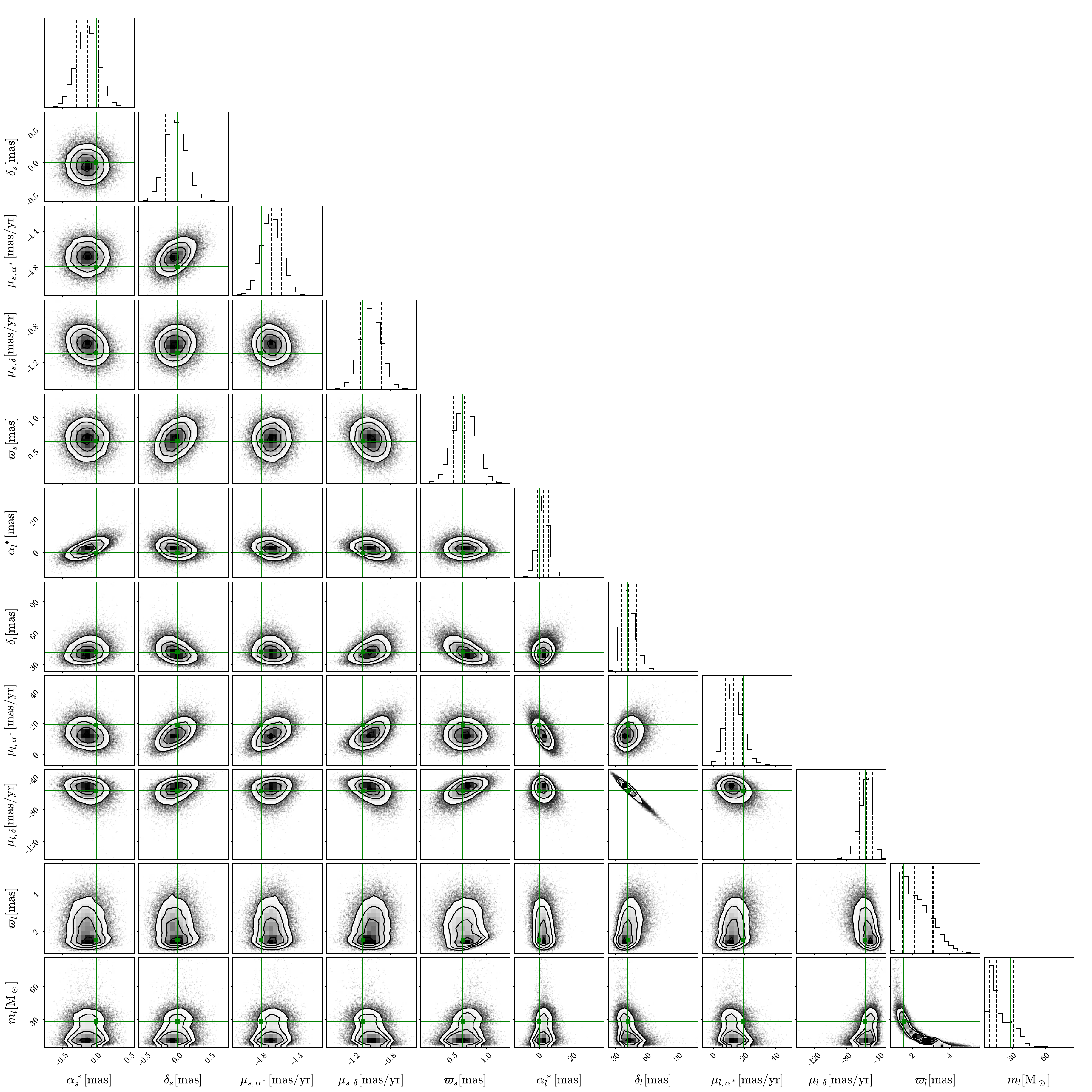}
    \caption{Corner plot of the blip model parameters fitted to source 5727504125199235456 via the constrained test statistic $\text{TS}^*$ in the BH mock catalog. The parameters are: The initial source displacement ($\alpha^*_s$,$\delta_s$), the source proper motion ($\mu_{s,a^*}$,$\mu_{s,\delta}$), the source distance $d_s$, the initial lens displacement ($\alpha^*_l$,$\delta_l$), the lens proper motion ($\mu_{l,a^*}$,$\mu_{l,\delta}$), the lens distance $d_l$, and the lens mass $m_l$. In green, the true parameter values. In dotted black, the 16th/50th/84th percentiles in the one-dimensional posteriors. \nblink{PaperPlots/corner_plot.ipynb}}
    \label{fig:covariance}
\end{figure}

We analyze the mock catalog described in section~\ref{sec:bh_catalog} to test the pipeline's ability to search for isolated astrophysical BHs in the MW. We conduct this blip search blindly. A total of 6 events pass the $5\sigma$ free model cut and our $\text{TS}^*>100$ requirement.  Out of those six, two do not pass the $3\sigma$ cut after the acceleration fit. Upon comparing with truth information (unblinding), we learn that all six of these events are true blips, demonstrating that the pipeline is capable of flagging astrophysical BHs in the \textit{Gaia} data and simultaneously not generating any false positives. These 6 events, their statistics, and their best fit parameters are shown in table~\ref{tab:blip_event_list}. Furthermore, the raw AL scan fits and residuals for two of these events are shown in figures~\ref{fig:sig1} and \ref{fig:sig2}; the remaining four plots are available on GitHub~\nblink{PaperPlots/significant_plot.ipynb}. Finally, figure~\ref{fig:covariance} shows the covariance between blip model parameters at the global maximum constrained log likelihood ratio (i.e.~where $\text{TS}^*$ is computed) for one of the six events; the other five corner plots are available at this link~\nblink{PaperPlots/corner_plot.ipynb}.

We also compute the source brightness magnification due to photometric lensing for each of these 6 events. Of the 4 that pass the $3\sigma$ acceleration fit cut, only one event has a magnification above the \textit{Gaia} photometric uncertainty. This demonstrates the advantage of looking for lensing signals with astrometric surveys. See appendix~\ref{sec:photometric} for a more detailed discussion and supplemental photometry plots.

\begin{table}[!htp]\centering
\begingroup
\renewcommand*{\arraystretch}{1.44}
\scriptsize
\begin{tabular}{c|ccccccc}\toprule
\\
ID &\multicolumn{2}{c}{\textbf{6664358989221213184}} &\multicolumn{2}{c}{\textbf{5727504125199235456}} &\multicolumn{2}{c}{\textbf{5931325238592343680}} \\
&\multicolumn{2}{c}{} &\multicolumn{2}{c}{} &\multicolumn{2}{c}{} \\
$d_s$ [pc] &\multicolumn{2}{c}{$1205.97$} &\multicolumn{2}{c}{$1531.68$} &\multicolumn{2}{c}{$2336.97$} \\
$\text{blippiness}$ &\multicolumn{2}{c}{$2.34$} &\multicolumn{2}{c}{$18.81$} &\multicolumn{2}{c}{$2.30$} \\
G mag &\multicolumn{2}{c}{$14.45$} &\multicolumn{2}{c}{$18.87$} &\multicolumn{2}{c}{$16.21$} \\
&\multicolumn{2}{c}{} &\multicolumn{2}{c}{} &\multicolumn{2}{c}{} \\
$-2\log\hat{\mathcal{L}}_{\rm free}$ &\multicolumn{2}{c}{$244.38$} &\multicolumn{2}{c}{$676.40$} &\multicolumn{2}{c}{$996.95$} \\
$-2\log\hat{\mathcal{L}}_{\rm accel}$ &\multicolumn{2}{c}{$92.26$} &\multicolumn{2}{c}{$478.30$} &\multicolumn{2}{c}{$133.44$} \\
$\text{TS}^*$ &\multicolumn{2}{c}{$173.26$} &\multicolumn{2}{c}{$593.91$} &\multicolumn{2}{c}{$938.33$} \\
$\text{TS}$ &\multicolumn{2}{c}{$173.78$} &\multicolumn{2}{c}{$594.42$} &\multicolumn{2}{c}{$940.67$} \\
\\
&\textbf{Best Fit} &\textbf{Truth} &\textbf{Best Fit} &\textbf{Truth} &\textbf{Best Fit} &\textbf{Truth} \\
$\Delta\alpha^*$ [mas] &$132.58^{+44.97}_{-49.52} $ &$122.56$ &$2.22^{+3.43}_{-3.32} $ &$-0.21$ &$-324.78^{+101.49}_{-113.00}$ &$-317.95$ \\
$\Delta\delta$ [mas] &$-35.39^{+52.72}_{-79.54}$ &$-56.05$ &$42.01^{+7.70}_{-5.92}$ &$41.861$ &$-48.40^{+38.41}_{-52.77}$ &$-121.17$ \\
$\mu_{\alpha*}$ [mas/yr] &$40.61^{+17.56}_{-16.45}$ &$39.84$ &$12.81^{+5.93}_{-5.02}$& $19.04$ &$23.11^{+25.70}_{-19.13}$ &$-7.36$ \\
$\mu_{\delta}$ [mas/yr] &$-0.55^{+22.38}_{-21.95}$ &$-5.54$ &$-54.65^{+7.38}_{-9.32}$ &$-57.14$ &$94.45^{+37.63}_{-31.17}$ &$123.07$ \\
$d$ [pc] &$588.23^{+218.22}_{-162.70}$ &$504.69$ &$467.29_{-145.75}^{+212.98} $ &$650.61$ &$184.84_{-42.79}^{+76.25}$ &$179.34$ \\
$m$ [$M_{\odot}$] &$9.89^{+8.47}_{-2.41}$ &$10.00$ &$15.97^{+15.11}_{-6.20}$ &$28.30$ &$33.32^{+5.69}_{-6.22} $ &$27.74$ \\
\\
\cline{0-7} \\
ID &\multicolumn{2}{c}{\textbf{6262458554071571712}} &\multicolumn{2}{c}{\textbf{4042201774850362496}} &\multicolumn{2}{c}{\textbf{4068042664558486272}} \\
&\multicolumn{2}{c}{} &\multicolumn{2}{c}{} &\multicolumn{2}{c}{} \\
$d_s$ [pc] &\multicolumn{2}{c}{$3709.04$} &\multicolumn{2}{c}{$6489.581$} &\multicolumn{2}{c}{$7184.126$} \\
$\text{blippiness}$ &\multicolumn{2}{c}{$6.99$} &\multicolumn{2}{c}{$8.510$} &\multicolumn{2}{c}{$3.35$} \\

G mag &\multicolumn{2}{c}{$18.06$} &\multicolumn{2}{c}{$18.51$} &\multicolumn{2}{c}{$16.79$} \\
&\multicolumn{2}{c}{} &\multicolumn{2}{c}{} &\multicolumn{2}{c}{} \\
$-2\log\hat{\mathcal{L}}_{\rm free}$ &\multicolumn{2}{c}{$208.61$} &\multicolumn{2}{c}{$217.24$} &\multicolumn{2}{c}{$1538.27$} \\
$-2\log\hat{\mathcal{L}}_{\rm accel}$ &\multicolumn{2}{c}{$94.107$} &\multicolumn{2}{c}{$166.93$} &\multicolumn{2}{c}{$451.82$} \\
$\text{TS}^*$ &\multicolumn{2}{c}{$158.25$} &\multicolumn{2}{c}{$147.86$} &\multicolumn{2}{c}{$1437.47$} \\
$\text{TS}$ &\multicolumn{2}{c}{$159.22$} &\multicolumn{2}{c}{$150.25$} &\multicolumn{2}{c}{$1439.82$} \\
\\
&\textbf{Best Fit} &\textbf{Truth} &\textbf{Best Fit} &\textbf{Truth} &\textbf{Best Fit} &\textbf{Truth} \\
$\Delta\alpha^*$ [mas] &$5.26^{+2.38}_{-2.33}$ &$4.30$ &$-90.42^{+61.59}_{-64.82}$ &$-51.78$ &$-199.17^{+68.82}_{-54.04}$ &$-196.75$ \\
$\Delta\delta$ [mas] &$13.21^{+7.25}_{-3.85}$ &$19.25$ &$42.85^{+36.26}_{-26.24}$ &$22.60$ &$-124.84^{+42.41}_{-43.99}$ &$-132.45$ \\
$\mu_{\alpha*}$ [mas/yr] &$12.66^{+6.03}_{-3.80}$ &$11.91$ &$7.92^{+29.80}_{-19.78}$ &$12.642$ &$-20.31^{+12.29}_{-15.51}$ &$-9.54$ \\
$\mu_{\delta}$ [mas/yr] &$3.61^{+3.92}_{-2.84}$ &$7.29$ &$-140.43^{+79.82}_{-90.61}$ &$-86.42  $ &$-116.11^{+37.21}_{-32.17}$ &$-114.30$ \\
$d$ [pc] &$2980.95^{+144.05}_{-1129.10}$ &$2604.89$ &$255.75^{+343.05}_{-101.19}$ &$439.65$ &$302.11_{-86.13}^{+156.61}$ &$360.32$ \\
$m$ [$M_{\odot}$] &$26.22^{+8.21}_{-6.01}$ &$33.56$ &$7.70^{+1.72}_{-1.15}$ &$7.640$ &$27.24^{+6.41}_{-6.46}$ &$31.90$ \\
\bottomrule
\end{tabular}
\endgroup
\caption{Overview of all six source trajectories in the BH mock catalog with $\text{TS}^*>100$, all of which are true blip events. We list the source ID, the true distance to each source $d_s$, the true blippiness of each event, the G magnitude of each source, the test statistics corresponding to three model fits (free, acceleration, and blip), and the best fit and true lens parameters, estimated based on nested sampling quantiles obtained via \texttt{PyMultinest}.}
\label{tab:blip_event_list}
\end{table}

One of these events (top row, second column in table~\ref{tab:blip_event_list}) has best fit values particularly close to the true lens parameters with narrow error bars. This is because the lens is rather close and has a high blippiness value. This event breaks much of the parameter degeneracy that plagues more distant and less significant blip trajectories. The lens distance degeneracy with mass and proper motion can also be seen in figure~\ref{fig:covariance}, and is much stronger for the other 5 blip events. These degeneracies explain why the parameters that maximize the constrained likelihoods do not necessarily coincide with the true lens parameters.

We conclude that we likely expect to see about four true blip events after both the acceleration and $\text{TS}^*$ cut, the closest and most blippy of which will have accurately determined lens parameters. The sources in question and candidate lens locations should then be followed up by other telescopes, providing exciting prospects for the study of phenomena associated with free-floating astrophysical BHs: e.g.~accretion from the interstellar medium~\cite{Fujita_1998,agol2002}, and superradiance~\cite{1971JETPL..14..180Z,PhysRevLett.28.994,Starobinsky:1973aij,lasenby_2016,Baryakhtar_2017,Baryakhtar_2021}. A free-floating astrophysical BH has only been claimed to have been detected once in the past~\cite{Sahu_2022,2022ApJ...933L..23L,2022ApJ...937L..24M}.

\subsection{Binary systems}
\label{sec:binaries}

The exact fraction of stars in binary or higher order systems in the MW has not been accurately estimated, but surveys of Sun-like stars in the solar neighborhood suggest that it may be approximately half of all stars~\cite{Raghavan_2010}. Binaries that are entirely or partially resolved have been studied extensively using \textit{Gaia} data~\cite{10.1093/mnras/stab323}, and many of these sources are automatically flagged in \textit{Gaia}'s public data releases. Therefore, we may simply discard them from our analysis of the full astrometric DR4 catalog. \textit{Gaia}'s DR3 binary flagging procedure is described in ref.~\cite{Halbwachs2022}. \textit{Gaia} does not flag sources in binary orbits with a dark (or faint) companion, such as a neutron star, a brown dwarf, an exoplanet, or an astrophysical BH. It is known that the binary orbits of these sources induce a measurable correction to the free trajectory of the source~\cite{Kervella_2019,Andrews_2019,https://doi.org/10.48550/arxiv.2110.05549,Belokurov_2020,Penoyre_2022}. In particular, it is estimated that \textit{Gaia} is capable of observing about $75$ sources with BH companions~\cite{https://doi.org/10.48550/arxiv.2110.05549}. \textit{Gaia}'s binary flagging system is also conservative since flawed binary flagging can hurt \textit{Gaia}'s science output --- it is projected that some fully luminous binaries will bypass Gaia's flagging procedure. Ref.~\cite{El_Badry_2022} describes Gaia BH1: a binary system consisting of a G-type star orbiting a BH. This event is well described by the \textit{Gaia} DR3 binary orbit astrometric solution. It is estimated that there are sources like Gaia BH1 that go unflagged as binaries in the full \textit{Gaia} database.

To test our pipeline's ability to distinguish between trajectories of sources with an unresolved binary companion and true blips, we follow a test procedure similar to the one carried out in ref.~\cite{Andrews_2019}; that is, we generate a mock catalog consisting of $10^3$ luminous stars with masses of either $1~M_{\odot}$ or $10~M_{\odot}$, each with dark companions with masses corresponding to brown dwarfs, white dwarfs, neutron stars, or BHs (0.05~$M_{\odot}$, 0.6~$M_{\odot}$, 1.4~$M_{\odot}$, and 10~$M_{\odot}$, respectively). Note that the light centroid of two luminous but unresolved stars will follow a trajectory similar to that of a star with a non-luminous companion, like the ones we sample over here. We place these companion objects at distances of 10~pc, 100~pc, and 1~kpc. For each of these combinations of masses and distances, we probe orbital periods of $10$, $10^2$, $10^3$, and $10^4$~days, with the binary eccentricity drawn from a uniform distribution ranging from 0 to 0.95 and orbital Euler angles drawn from a uniform distribution ranging from 0 to $2\pi$. We then fit our free model, acceleration model, and blip model to the resultant stellar trajectories.

\begin{figure}
    \centering
    \includegraphics[width=\textwidth,height=\textheight,keepaspectratio]{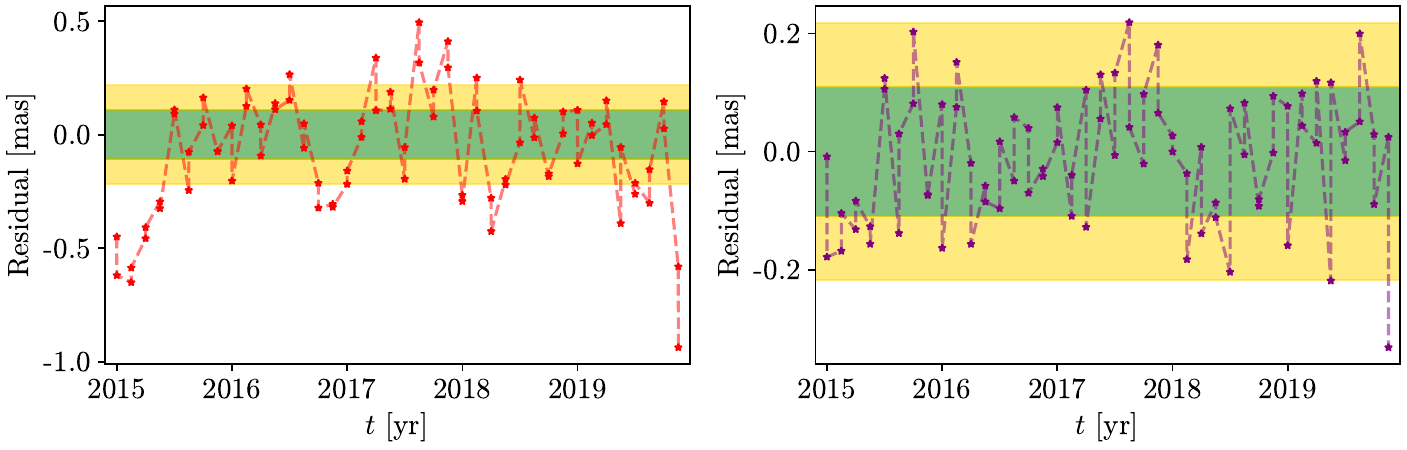}
    \caption{Residuals for free and acceleration models fitted to the trajectory of a source with a BH binary companion and orbital period $t_{\rm bin}=10^4$ days. \textit{Left:} In red, the residual from the free model fit to the source trajectory, with the 1$\sigma$ and 2$\sigma$ bands being shown in green and yellow, respectively. \textit{Right:} In purple, the residual from the acceleration model fit to the source trajectory. Note that the dashed lines do not show the continuous trajectory of the model and data, but rather simply connect the data points since their order can otherwise be hard to gauge. The free fit exceeds the $5\sigma$ free log likelihood cutoff; however, it is far below $3\sigma$ significance in the acceleration fit, meaning it fails to qualify as a blip (even without accounting for its associated constrained test statistic $\text{TS}^*$. The acceleration fit and cut effectively eliminates sources that are part of long-period binaries. \nblink{PaperPlots/binary_accel_plot.ipynb}}
    \label{fig:binary_companion}
\end{figure}

We find that there are two classes of binaries, depending on which cuts are passed and which are failed. The first type of binary has an orbital period $t_{\text{bin}}$ longer than \textit{Gaia}'s observation time ($t_{\text{bin}}\gg t_{\text{obs}}$). These binaries can have significant free model log likelihoods, but their significance becomes much smaller when fit to the acceleration model due to their trajectory being well approximated by a star undergoing constant angular acceleration in a single direction. In our grid catalog, all of these binaries have acceleration fit log likelihoods below the $3\sigma$ interest threshold $-2\log\hat{\mathcal{L}}_{\rm accel}<\chi^2_{3\sigma}$. See figure~\ref{fig:binary_companion} for an example of a fit of this type.

The second type of binary has a period comparable to or smaller than the observation time ($t_{\rm obs}\gtrsim t_{\rm bin}$) and typically has a significant free log likelihood $-2\log\hat{\mathcal{L}}_{\rm free}>\chi^2_{5\sigma}$, as well as a significant acceleration log likelihood $-2\log\hat{\mathcal{L}}_{\rm accel}>\chi^2_{3\sigma}$. However, because blips and short period binaries have very distinct trajectories, for most of the sources, $\text{TS}^*<100$. Binary trajectories are also disfavored by the priors we use to constrain the computation of $\text{TS}^*$. However, for two sources in the catalog, even this cut is surpassed. To avoid accidentally flagging sources with dark companions as blips, we thus impose the cut $\text{TS}^*<\text{TS}_{3\sigma}$, where the $\text{TS}_{3\sigma}$ is the $3\sigma$ significance threshold for the blip model \textit{unconstrained} $\text{TS}$ distribution obtained via MC generated blip events. All six of the blip events in section~\ref{sec:bh_catalog} pass this cut. The actual number of dark companion that \textit{Gaia} expects to see is much smaller than the number we have considered here, so it is likely that this extra cut is unnecessary. We nevertheless implement it into the analysis pipeline as a precautionary measure.
Finally, we note that binaries that are completely dark, e.g.~consisting of two black holes, are indistinguishable from isolated black holes in the sky when the binary angular separation is much smaller than the Einstein radius. These dark binaries have never been directly observed and are thus another interesting lens population to probe.

\subsection{Projected compact DM constraints}
\label{sec:PBH_constr}
\begin{figure}
    \centering
    \includegraphics[width = \textwidth]{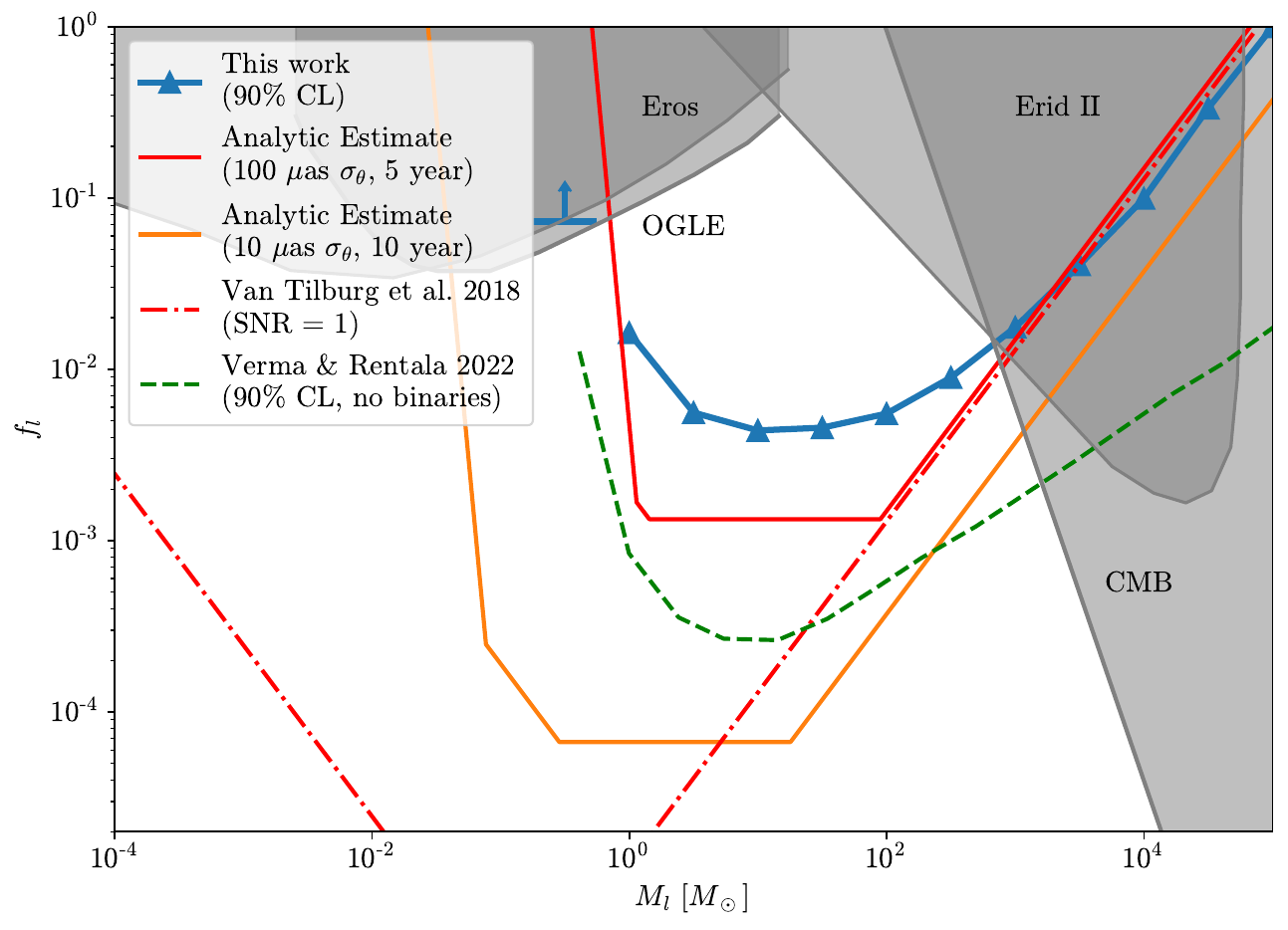}
    \caption{The blue curve is the projected 90\% constraint of DM fraction $f_l$ in the form of compact objects from the analysis in this work, assuming no other astrophysical backgrounds. Our sensitivity is peaked around $10$--$100~M_\odot$ and sharply evaporates below $1~M_\odot$ because there the Einstein radius is smaller than the astrometric precision of \textit{Gaia}. At larger masses, the sensitivity to $f_l$ decreases linearly due to the decrease in lens number density at fixed $f_l$.  We overlay the analytic $\text{SNR} = 1$ estimate (dot-dashed red curve) of ref.~\cite{van2018halometry} and our updated analytic estimate for the 90\%-CL exclusion limit for a blind astrometry-only analysis (solid red). The reach of a hypothetical 10-year future mission with $10\times$ better astrometric uncertainties is shown in the orange solid curve using the same analytic estimates. The 90\%-CL exclusion curve using the probabilistic model of ref.~\cite{Verma} is depicted as the green dashed curve. 
    Existing constraints from MW photometric microlensing \cite{EROS,wyrzykowski2011ogle2}, dwarf galaxy heating \cite{ERID2}, and CMB spectral distortions from PBH accretion \cite{CMB_PBH} are shown by gray shaded regions. \nblink{PaperPlots/PBH_limit.ipynb}}
    \label{fig:PBH}
\end{figure}

Figure~\ref{fig:PBH} shows the projected constraining power of \textit{Gaia} DR4 on compact DM, following the procedure of section~\ref{sec:constrain_method}. To arrive at this result, we inject $7.0\times10^9$ compact DM objects into the mock catalog (corresponding to 10\% of total DM mass for $1 ~ M_\odot$ compact objects). The blue curve shows the resulting 90\%-CL limits on mock simulations with delta-function compact DM object mass functions over the range $10^{-1}$--$10^5~M_\odot$. For compact objects lighter than $0.3~M_\odot$, there is no event in the signal region, so we are only able to quote an upper bound on $f_l$ as shown by the blue arrow. The sensitivity peaks at compact object masses between $1~M_\odot$ and $100~M_\odot$. At smaller masses, the sensitivity sharply decreases due to the saturation of astrometric deflection at the Einstein radius, while for larger masses it decreases more gradually due to the smaller expected number of compact objects with a large blippiness. Existing constraints from photometric microlensing~\cite{EROS,wyrzykowski2011ogle2}, dwarf galaxy heating~\cite{ERID2}, and CMB spectral distortions (from X-ray accretion onto PBHs, not applicable for non-PBH compact objects)~\cite{CMB_PBH} are shown in gray.

We also show in figure~\ref{fig:PBH} the initial analytic estimate from ref.~\cite{van2018halometry} for the potentially accessible parameter space of compact DM objects (red dot-dashed curve). At the low-mass end, their estimate is a contour for which the \emph{local} signal-to-noise ratio equals unity. Without any additional input from other surveys to identify potential astrometric lensing candidates, the look-elsewhere effect and the requirement of setting a 90\%-CL limit drastically reduces the projected constraints on the DM fraction at low masses, equivalent to setting $\text{SNR} = 15$ in the language of ref.~\cite{van2018halometry}. The requirement of such a high threshold for a blind search furthermore means that the weak lensing approximation no longer holds, further suppressing the sensitivity of a blind search purely based on astrometry alone.
In appendix~\ref{sec:pbh_estimate}, we recalculate the analytic estimate following the same procedure in ref.~\cite{van2018halometry} with the above-mentioned effects and arrive at the updated analytic estimate shown in the red solid curve, which is much closer to the mock catalog simulation. We also show the projected reach of a futuristic 10-year mission with astrometric uncertainties 10 times better than the \textit{Gaia} EDR3 uncertainties in solid orange. The contour of 2.3 detectable events (corresponding to a 90\% constraint) from ref.~\cite{Verma} is shown by the dashed green curve. The difference between our work and that of ref.~\cite{Verma} can also be partially ascribed to differences in treatment of the look-elsewhere effect. The scaling difference at large compact object masses is because that we conservatively discard events that have an acceptable (within $3\sigma$) 7-parameter acceleration fit, necessary to eliminate backgrounds from long-period binary systems.

\section{Conclusions}
\label{sec:conclusions}

Precision astrometric measurements from $\textit{Gaia}$ enable a new way to probe the MW for transient astrometric lensing caused by massive non-luminous objects of either astrophysical or primordial origin, with potential for discovering several free-floating BHs and searching for compact objects down to a very small fraction of DM. We construct an analysis pipeline (\githubmaster) capable of systematically and exhaustively searching for transient astrometric lensing events (or ``blips'') in the upcoming $\textit{Gaia}$ DR4 catalog. This pipeline works by first fitting a simple free (unlensed) model of stellar motion to more than a billion stars in DR4 using a combination of traditional optimization and bayesian inference. It then discards all events that are not more than $5\sigma$ outliers under the free stellar motion hypothesis. To account for binaries, the pipeline then fits a model of stellar motion in which the source being studied undergoes constant angular acceleration. Events that are not more than $3\sigma$ outliers under this constant angular acceleration hypothesis are similarly discarded. Finally, the pipeline fits a blip model, weighted by priors on lens proper motion, distance, and mass, to the remaining events. Any events that pass the free fit and acceleration fit cuts and that have blip test statistics $\text{TS}^* > 100$ are flagged as blip candidates. Using the Yellin method, the pipeline furthermore infers constraints on dark compact object populations based on the test statistic distribution.

To test the pipeline, we create three types of mock DR4 catalogs based on the currently available EDR3 catalog. The first contains no dark lenses, meaning all sources undergo free stellar motion. In this catalog, the pipeline flags no events as being blips, and the log likelihood distribution follows the $\chi^2$ expectation (see figure~\ref{fig:x0_chisq}). The second mock catalog is identical to the first, except we inject astrophysical BHs based on current priors on the BH number density and proper motion distribution across the MW. In this catalog, we find 4 lensing events that pass all of our cuts; namely, they are above $5\sigma$ significance under the free model expectation and above $3\sigma$ significance under the acceleration model fit, which separates the events from long-period binary systems with a dark companion, and they have a constrained test statistic $\text{TS}^*>100$. This gives us a benchmark of the total number of astrometric lensing events by isolated astrophysical BHs we expect to discover in \textit{Gaia} DR4.

We inject the third mock catalog with compact objects of a single mass spanning the range $10^{-1}$--$10^5~M_\odot$ to constrain their fraction of DM in the MW using the Yellin method. Our projected constraint indicates that \emph{Gaia} has leading reach on the compact DM fraction in the mass range of $1$--$10^3~M_\odot$. We find that \textit{Gaia} loses sensitivity for point-like DM lenses lighter than $0.3~M_\odot$, is most sensitive between $10$--$100~M_\odot$ (projected exclusion fraction of $f_l\sim4\times10^{-3}$), and runs out of observable blip events for higher masses as the number of lenses and thus transient lensing events decreases. Our full \textit{Gaia} DR4 mock catalog enables us to properly assess the statistical background of the large data set to obtain faithful projections of discovery potential and constraints.

We make a few assumptions and simplifications in creating the mock \textit{Gaia} DR4 catalog which will be different from the actual \textit{Gaia} DR4. Here we outline those points and the potential effect on the actual data analysis with real \textit{Gaia} data.
\begin{itemize}
    \item We assume all sources in \textit{Gaia} will be observed exactly 80 times, roughly the sky-averaged expected number of observations. This is not the case for the real data. Each source will be observed roughly $60$--$140$ times depending on the the source's ecliptic latitude. If the high-cadence region has a larger/smaller overlap with the region of higher stellar density (e.g.~Galactic plane), then we would expect more/fewer lensing events discovered compared to the mock catalog.
    \item We assume \textit{Gaia} only records the one-dimensional offset along the AL direction for all stars. This is not true for the brightest stars. They will have the full two-dimensional trajectory in the AL and AC direction recorded. However, the uncertainty in the AC direction is orders of magnitude worse than that of the AL direction due to design of the telescope. This will only improve sensitivity of the brightest stars by a small margin.
    \item The AL uncertainties we adopt in the mock catalog are the projected optimal uncertainties of DR4 reported in \textit{Gaia} EDR3. If the actual uncertainties are different, the sensitivity projections in this work will be affected accordingly.
\item We only inject astrophysical BHs for our search for compact objects. In reality, there will be other compact objects, such as neutron stars, white dwarfs, brown dwarfs, and faint main sequence stars. These objects could affect our projection, although we argue that their effect will be marginal (see appendix~\ref{sec:other_compact}). Potential contamination due to these other sources must nevertheless be carefully accounted for when DR4 is released and real data is available.
    \item We only use the effects of astrometric lensing for finding compact lens in this work. \textit{Gaia} DR4 will also release time-series photometric measurements of the stars. Although \textit{Gaia}'s photometric capabilities are not optimal for lensing searches, a combination of its photometric and astrometric measurements will likely lead to more precise lens parameters and potentially stronger discovery potential, especially for low-mass lenses for which strong lensing events are more common.
\end{itemize}

Beyond the single-source blip search outlined here, it is also interesting to consider events in which a non-luminous lens affects the astrometric trajectory of \emph{multiple sources} in a short time interval. Such events may not be detectable by probing for solitary blips, since the lensing deflection of any given source might be too small to be statistically significant. Furthermore, observing two or more sources undergoing gravitational lensing due to the same lens would likely yield a much better determination of the physical parameters of the lens. Conventional likelihood optimization, as used in this work, is likely not computationally feasible for carrying out a ``multi-blip'' search due to the number of free parameters in such a model. Machine learning tools will likely accelerate the pattern recognition of those correlated lensing deflections---an avenue we will explore in future work.

Our analysis pipeline and mock catalog are not just applicable to \textit{Gaia} DR4. The tools we provide in this work can be used on past astrometry legacy archives (e.g. HSTPROMO~\cite{van2013local}, PHAT~\cite{dalcanton2012panchromatic}), as well as future astrometric surveys (e.g.~the Nancy Grace Roman Space Telescope (formerly known as WFIRST)~\cite{Roman}, GaiaNIR~\cite{hobbs2017gaianir,hobbs2019voyage}, THEIA~\cite{THEIA} with minor adjustments, and of course \textit{Gaia} DR5, which is projected to contain all collected \textit{Gaia} data~\cite{ESA_data_release}). Charting out several isolated, electromagnetically quiet BHs will be a major milestone in astrophysics, and help in the understanding of their formation mechanisms. Finally, isolated BHs are also pristine laboratories for Beyond the Standard Model Physics searches. The extreme gravity near a BH can give rise to BSM signals, most notably through superradiance~\cite{1971JETPL..14..180Z,PhysRevLett.28.994,Starobinsky:1973aij,lasenby_2016,Baryakhtar_2017,Baryakhtar_2021}.

Transient astrometric weak lensing is a powerful probe of the distribution and properties of known compact remnants, such as BHs and neutron stars, as well as extreme overdensities in the DM distribution. We look forward to the application of our tools to these studies.

\acknowledgments

We thank Vasily Belokurov, Anthony Brown, Kyle Cranmer, Neal Dalal, Joshua W. Foster, David Hogg, Jessica Lu, Peter McGill, Siddharth Mishra-Sharma, and Neal Weiner for several insights and discussions, and Cyril Creque-Sarbinowski, David Dunsky, Cara Giovanetti, Siddharth Mishra-Sharma, and Andreas Tsantilas for helpful comments on the manuscript. We also thank the referee, {\L}ukasz Wyrzykowski, for constructive feedback.
This material is based upon work supported by the National Science Foundation under Grant No.~2210551.
The authors are grateful for the hospitality of Perimeter Institute, where part of this work was performed.
This work was supported in part through the NYU IT High Performance Computing resources, services, and staff expertise. The Center for Computational Astrophysics at the Flatiron Institute is supported by the Simons Foundation. Research at Perimeter Institute is supported in part by the Government of Canada through the Department of Innovation, Science and Economic Development Canada and by the Province of Ontario through the Ministry of Colleges and Universities. This work has made use of data from the European Space Agency (ESA) mission {\it Gaia} (\url{https://www.cosmos.esa.int/gaia}), processed by the {\it Gaia} Data Processing and Analysis Consortium (DPAC, \url{https://www.cosmos.esa.int/web/gaia/dpac/consortium}). Funding for the DPAC has been provided by national institutions, in particular the institutions participating in the {\it Gaia} Multilateral Agreement. We have made use of the software packages \texttt{PyMultinest}~\cite{Feroz_2008,Feroz_2009,Feroz_2019,buchner}, \texttt{corner}~\cite{corner}, \texttt{Astropy}~\cite{astropy:2013,astropy:2018,astropy:2022}, \texttt{SciPy}~\cite{scipy}, \texttt{healpy}~\cite{healpy}, \texttt{HEALPix}\footnote{\url{http://healpix.sourceforge.net}}~\cite{HEALPix}, and \texttt{NumPy}~\cite{numpy}.

\bibliographystyle{unsrt}
\bibliography{main}

\appendix

\section{Extended objects}
\label{sec:extended}
Extended objects, such as DM subhalos, are also potential targets for transient astrometric lensing searches.
However, we will show in this section that the blip technique demonstrated in this paper is not sensitive to astrometric lensing caused by a gravitationally collapsed MW subhalo in a standard cosmology.

\begin{figure}
    \centering
    \includegraphics[width = \textwidth]{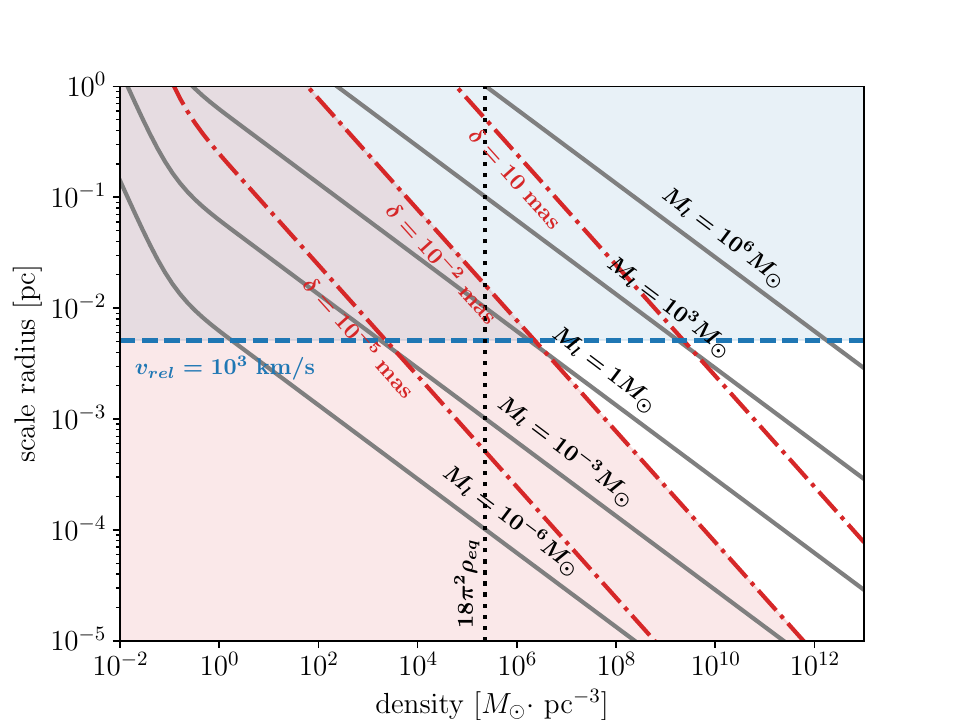}
    \caption{This figure shows the detectable region for an extended lens following a Gaussian density profile eq.~\eqref{eq:gaussian}. The gray solid line shows the contour of constant lens mass with the turning point being the Roche radius at 8 kpc. The red dashed-dotted line shows the maximum deflection an extended lens can cause (which equates to the minimum impact parameter being the scale radius). We exclude the parameter space of $\delta_\text{max} < 10~\mu$as which is our fiducial value of the \textit{Gaia} DR4 sensitivity. The blue dashed line and the shaded region show the blip requirement for a source-lens relative velocity of $10^3~$km/s, which corresponds to the lens and and source both moving at the galactic escape velocity back-to-back. The vertical black dotted line shows the density of a subhalo that gravitationally collapsed at matter-radiation equality. \nblink{PaperPlots/extended_subhalo.ipynb}}
    \label{fig:extend}
\end{figure}

For simplicity of calculation, we assume the DM subhalo has a Gaussian density profile given as
\begin{equation}
    \rho_l(r) = \frac{M_l}{4\pi}\frac{\exp{\frac{-r^2}{2r_l^2}}}{rr_l^2},
    \label{eq:gaussian}
\end{equation}
where $M_l$ is the mass of the lens and $r_l$ is the scale radius of the lens. We define the mean lens density as $\rho_{l,0} = M_l/r_l^3$. The resultant relation between $r_l$ and $\rho_{l,0}$ is shown in figure~\ref{fig:extend}. The solid line is the contour of total lens mass. The turning point near large scale radius is where the scale radius is equal to the Roche radius of the MW at 8 kpc. The dashed-dotted line is the contour of the maximum deflection an extended lens can induce. We require the deflection to be larger than $10~\mathrm{\mu as}$ to be detected by \textit{Gaia}, so extended lenses in the red-shaded region are not detectable. Another criterion for detection is eq.~\eqref{eq:blippiness}. Given that the maximum deflection of a extended lens occurs at the scale radius, the blip criterion is $v_\text{rel}\tau\geq r_l$. In the most conservative case where $v_\text{rel} = 1000~\mathrm{km/s}$, which corresponds to 2 objects moving back-to-back both at the galactic escape velocity, the requirement on $r_l$ is shown as the horizontal blue-dashed line.

The density of a subhalo that collapses at matter-radiation equality assuming a pure $\Lambda$CDM cosmology is shown as the vertical dotted line, marking the maximum density of a gravitationally collapsed subhalo in a standard cosmology. Since it is not within the range of the detectable parameter space, we only include lensing from point sources in this work.

\section{Other compact lens populations}
\label{sec:other_compact}
The analysis pipeline we present in the main text is also suitable for carrying out a blip search on compact lens populations other than BHs. Here, we provide preliminary estimates of the blip discovery potential of neutron stars, white dwarfs, brown dwarfs, and faint main sequence stars in \textit{Gaia} DR4.

\subsection{Neutron stars}
\label{sec:NS}
The MW is estimated to contain $10^8$--$10^9$ neutron stars~\cite{NS_num}, which is $1$--$10$ times the total number of BHs we inject in our mock catalog. The mass distribution of neutron stars is believed to lie within $1.0$--$2.2~M_\odot$, peaking at $1.4~M_\odot$~\cite{NS_mass}. Observations of neutron stars suggest that they, like BHs, receive natal kicks from supernovae, explaining their high velocities and large fractional abundance in the stellar halo~\cite{NS_num}. Therefore, we assume that the spatial and velocity distribution of neutron stars follow that of BHs.

We may thus use our mock analysis of astrophysical BH lensing from section~\ref{sec:results} to extrapolate the expected number of neutron star lensing events we will see in \textit{Gaia} DR4. Astrophysical BHs typically have a mass of about $10~M_\odot$ and neutron stars typically have a mass of around $1~M_\odot$. From figure~\ref{fig:PBH}, we see that the sensitivity from $10~M_\odot$ to $1~M_\odot$ drops by a factor of $\sim 5$. We assume there are $10^9$ neutron stars in the MW. Extrapolating, this means that the number of neutron star lensing events in DR4 with a significance level above $5\sigma$ is approximately one.

\subsection{White dwarfs}
The MW is estimated to contain approximately $10^{10}$ white dwarfs~\cite{WD_num}, which is $100$ times the total number of BHs contained in our mock catalog. The mass distribution of white dwarfs covers a range of $0.4$--$1.4~M_\odot$, peaking at $0.7~M_\odot$~\cite{WD_mass}. Observations of white dwarfs combined with simulations suggest that white dwarfs can be categorized into three families based on their kinematics: thin disk, thick disk, and halo~\cite{WD_num}. Each of the three families consist of $\mathcal{O}(1)$ of the total number of white dwarfs in the MW. To understand the discovery potential of isolated, faint white dwarfs in DR4, we follow the same procedure as in section~\ref{sec:NS}. Namely, we use the results of our BH mock analysis to extrapolate. Figure~\ref{fig:PBH} shows that the sensitivity drops sharply for lenses with a mass less than $1~M_\odot$. Thus, we are only sensitive to white dwarfs with a mass greater than $1~M_\odot$. Ref.~\cite{WD_mass} suggests that roughly $10\%$ of white dwarfs fall beneath this mass cutoff. So with $10^{10}$ white dwarfs in the MW, the number of potential observable white dwarfs is $10^{9}$. This suggests that the number of white dwarf blip events in DR4 with a significance level greater than $5\sigma$ is approximately one.

\subsection{Brown dwarfs}
Brown dwarfs are stellar objects with masses in the range $13~M_J$--$80~M_J~(1.2$--$7.6\times 10^{-2}~M_\odot)$, where $M_J$ is the mass of Jupiter. This is the mass range in which a star burns deuterium and hydrogen. Using the projected compact DM constraint shown in figure~\ref{fig:PBH}, we can see that the mass of a typical brown dwarf lies below \textit{Gaia}'s detectable range. This suggests that we will not see blip events caused by any isolated brown dwarfs in \textit{Gaia} DR4. One can also see this by using the analytic SNR estimate described by eq.~\eqref{eq:strong}. For a brown dwarf with a mass of $5\times10^{-2}~M_\odot$ located 10 (100, 1000) pc from the Sun, the maximum SNR one can get from astrometric lensing is 7 (4,  2), which is smaller than the the $\mathrm{SNR}=15$ threshold.
Thus, photometric microlensing is more suitable for the detection of brown dwarfs, cfr.~the shaded gray region of figure~\ref{fig:PBH}.

\subsection{Faint main sequence stars}
Main sequence (MS) stars are another possible lens population. A MS star passing in front of a background star can cause a blip event. Refs.~\cite{Kl_ter_2020,kluter2022prediction} discuss signals of star-star lensing and how to detect them in \textit{Gaia} data when both the lens and the background star are above \textit{Gaia}'s photometric threshold. They propose that star-star lensing can be used to determine the mass of luminous foreground stars. Here, we discuss blip events caused by faint MS stars dimmer than the \textit{Gaia} photometric threshold ($G \approx 21$).

As discussed in section~\ref{sec:bh_catalog}, all $5\sigma$ stellar BH events are within 1~kpc. This is due to the blippiness requirement described by eq.~\eqref{eq:blippiness}, which gives preference to lenses with large proper motions typically located at small line-of-sight distances. A MS star 1~kpc away from the Sun with an apparent magnitude of 20 will have an absolute magnitude of 10. Using the mass-luminosity relation
\begin{equation}
    \frac{L}{L_\odot} = \left(\frac{M}{M_\odot}\right)^{3.5},
\end{equation}
we estimate that the mass of such a star is roughly $\sim0.3~M_\odot$. Any MS star closer than 1~kpc that is too faint for \textit{Gaia} to detect must be lighter than this, which places the star outside \textit{Gaia}'s projected blip sensitivity shown in figure~\ref{fig:PBH}. Therefore, our preliminary analysis using mock catalogs suggests that \textit{Gaia} is not capable of discovering blips caused by faint MS stars.

\section{Results using GOST scanning law}
\label{sec:gost}

\begin{figure}
    \centering
    \includegraphics[width = 0.8\textwidth]{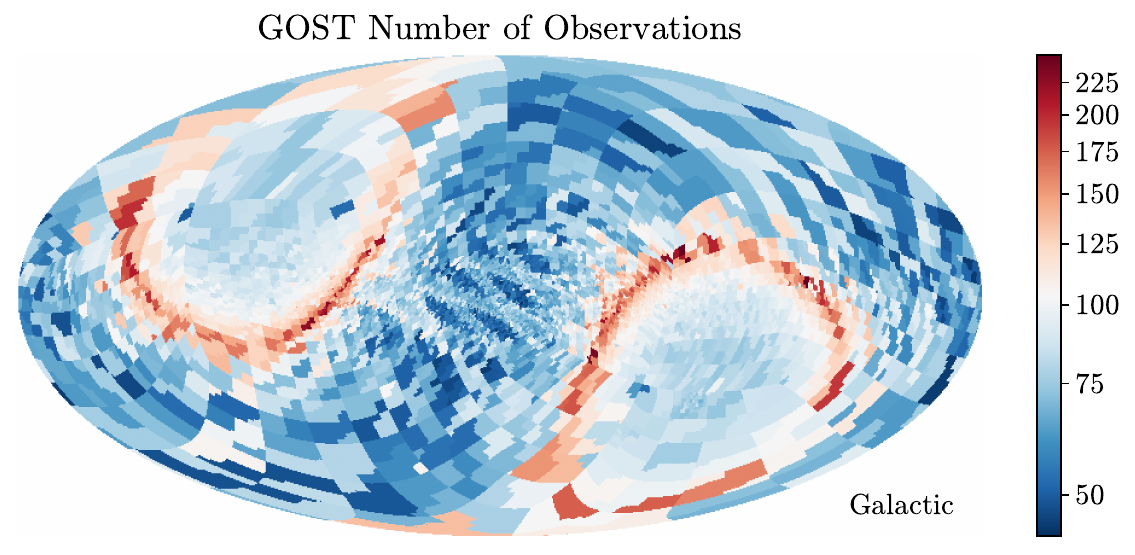}
    \caption{Location dependence of the number of observations obtained from GOST between Jan 2015 to Dec 2019. \nblink{PaperPlots/healpy_gost.ipynb}}
    \label{fig:gost_cadence}
\end{figure}

\begin{figure}
    \centering
    \includegraphics[width = \textwidth]{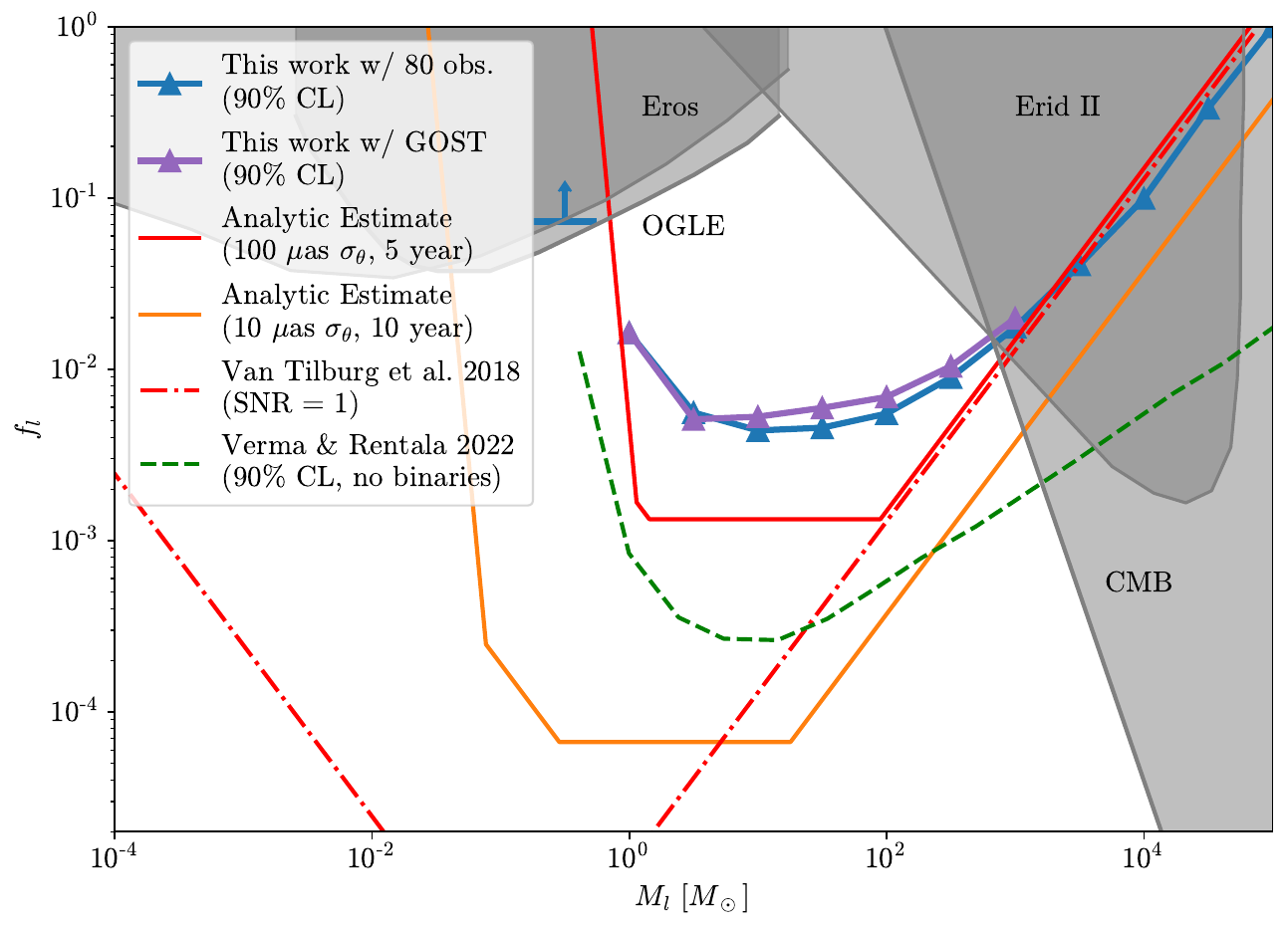}
    \caption{Compact DM constraint comparing the limits obtained from the updated GOST scanning law to the limits obtained from  the averaged 80 observations assumed in the main text. \nblink{PaperPlots/PBH_limit_gost.ipynb}}
    \label{fig:gost_PBH}
\end{figure}

Here, we discuss how using \textit{Gaia}'s Observation Forecast Tool (GOST) affects the results presented in section~\ref{sec:results}. To obtain accurate time-series data points for each \textit{Gaia} source, we compute the average angular location of each local batch of sources (with each batch corresponding to one of the 3386 \textit{Gaia} EDR3 files) by taking an average of their HealPIX location. Inputting this into GOST, we obtain the scanning law associated with each source. Using this method, the number of observations per source ranges from 43 to 249. The location dependence of \textit{Gaia}'s observation cadence is shown in figure~\ref{fig:gost_cadence}. We rerun the analysis described in section~\ref{sec:analysis} on mock catalogs generated using GOST. Limits obtained from the GOST DM mock catalog are shown in figure~\ref{fig:gost_PBH}. We note that these limits are marginally weaker than those obtained using the 80 data points scenario. Furthermore, we also conduct a BH search on a BH GOST mock catalog and find exactly 3 (6) highly significant sources with (without) the acceleration test statistic cut. This result corresponds almost exactly to the one obtained using the 80 data points method presented in the main text.

\section{Derivation of BH proper motion prior}
\label{sec:bh_pm_pdf}
Starting with the thin disk stellar distribution in eq.~\eqref{eq:thin_disk}, we can estimate the increase in $z_d$ by considering the following. We assume all stars start at exactly $z = 0$ with some known velocity dispersion $\sigma_{v_z}$. The probability distribution function (PDF) of stars at $z = 0$ is
\begin{equation}
    P(v_z) \propto \exp\left(-\frac{v_z^2}{2\sigma_{v_z}^2}\right).
\end{equation}
From energy conservation, the PDF of stars at $z$ is
\begin{equation}
    P(v_z) \propto \exp\left(-\frac{v_z^2}{2\sigma_{v_z}^2} - \frac{\phi(z)}{\sigma_{v_z}^2}\right),
\end{equation}
where $\phi(z)$ is the gravitational potential at $z$. Marginalizing over velocities gives:
\begin{equation}
    \frac{n(R, z)}{n(R,0)} = \exp\left(- \frac{\phi(z)}{\sigma_{v_z}^2}\right) = \exp\left(-\frac{|z|}{z_d}\right),
\end{equation}
where the second equals sign comes from eq.~\eqref{eq:thin_disk}. Here we can see that if the background gravitational potential stays the same, the scale height $z_d\propto\sigma_{v_z}^2$.

BH X-ray binaries (figure~7 in ref.~\cite{Atri_2019}) suggest a bimodal distribution of natal kick velocities. In \textit{Gaia} DR2, the vertical velocity dispersion around the solar neighborhood is reported to be around $\sigma_{v_z}\approx 20$ km/s \cite{2018}. Combining the stellar velocity dispersion and natal kick, the final velocity dispersion is approximately $\sigma_{v_z}\approx 70$ km/s. In terms of the scale height of the thin disk distribution, this implies that the scale height of BH distribution is around 10 times that of the scale height of stellar distribution. Therefore, we use $z_d = 3$ kpc for the BH distribution in the sky.

\begin{figure}
    \centering
    \includegraphics[width = .7\columnwidth]{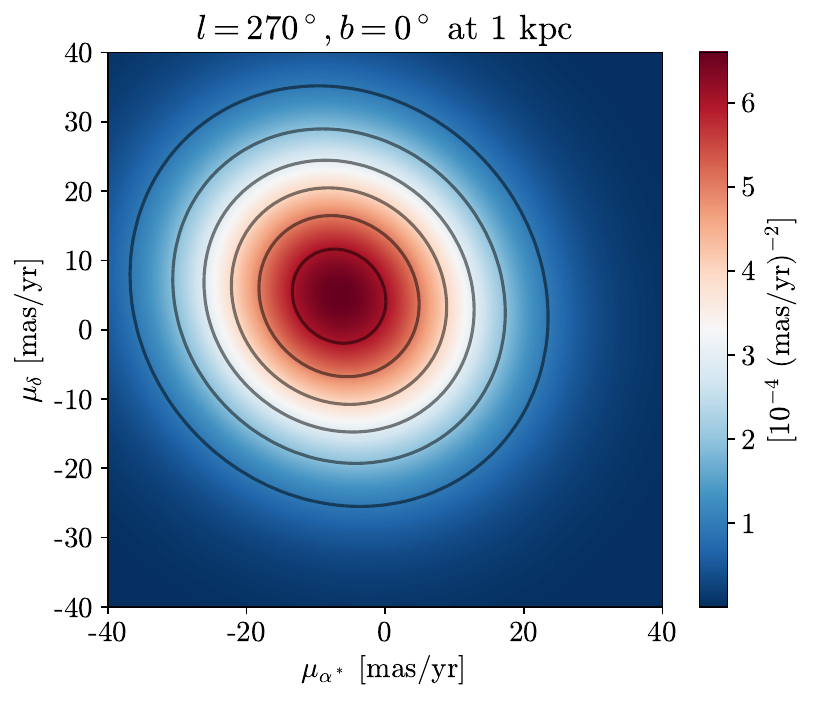}
    \caption{The conditional PDF of BH proper motion at $(l,b) = (270^\circ, 0^\circ)$ and $D_l = 1$ kpc. Here we can see the offset from the rotational velocity of the BH and the Sun. The effect from a non-diagonal velocity dispersion is also visible. The velocity dispersion $\pmb{\Sigma} = \text{diag}(77.5, 72.5, 70.0)$ km/s is again obtained by combining both the stellar velocity dispersion from ref.~\cite{2018} and the bimodal distribution of BH natal kicks from ref.~\cite{Atri_2019}. \nblink{PaperPlots/3d_prior.ipynb}}
    \label{fig:pm_dist_pdf}
\end{figure}

At a given location in galactic coordinate $(l,b)$, the joint distribution of the lens proper motion and distance $P(\pmb{\mu}_l, D_l|l,b)$ is given by Bayes' theorem
\begin{equation}
    P(\pmb{\mu}_l, D_l|l,b) = P(\pmb{\mu}_l | D_l, l, b)P(D_l|l, b).
\end{equation}
The distance prior $P(D_l|l, b)$ is given by eq.~\eqref{eq:distance_prior}. The conditional probability $P(\pmb{\mu}_l | D_l, l, b)$ can be calculated via the following process: we start with the conditional probability
\begin{equation}
\begin{split}
    P_\text{BH}(\mathbf{v}^C|D_l, l, b) &= \frac{1}{(2\pi)^{3/2}\det\pmb{\Sigma}}\exp\left(-\frac{1}{2}{\mathbf{v}^C}^T\pmb{\Sigma}^{-2}\mathbf{v}^C\right),\\
    \mathbf{v}^C &= \begin{pmatrix}
                v_R\\
                v_\phi\\
                v_z
                \end{pmatrix} -
                \begin{pmatrix}
                0\\
                -220\\
                0
                \end{pmatrix} \text{km/s},\\
    \pmb{\Sigma} &= \text{diag}(\sigma_{v_R}, \sigma_{v_\phi}, \sigma_{v_z}),
\end{split}
\end{equation}
where $\mathbf{v}^C$ is the linear velocity vector in a cylindrical coordinate centered at the galactic center and with $\phi = \pi$ pointing towards the solar system. $\pmb{\Sigma}$ is the velocity dispersion of the lens and we assume it is diagonal in this coordinate system. Next, we can rotate this into a Cartesian coordinate $(U,V,W)$ commonly used in astronomy where the galactic center sits at $(0,0,0)$, the solar system sits at $(-8,0,0)$~kpc, the $V$ axis points towards the direction of the Sun's orbit around the galactic center, and the $W$ axis points towards the galactic north pole. And, shift into a frame where the Sun is stationary. Then, the joint PDF in the Cartesian coordinate is
\begin{equation}
\begin{split}
    P_\text{BH}(\mathbf{v}^\mathbb{R}|D_l, l, b) &= \frac{1}{(2\pi)^{3/2}\det\pmb{\Sigma}}\exp\left(-\frac{1}{2}{\mathbf{v}^\mathbb{R}}^T\mathbf{R}_1\pmb{\Sigma}^{-2}\mathbf{R}_1^T\mathbf{v}^\mathbb{R}\right)\\
    \mathbf{v}^\mathbb{R} &= \mathbf{R}_1\mathbf{v}^C - \mathbf{v}_\odot^\mathbb{R}\\
    \mathbf{R}_1 &= \begin{pmatrix}
                    \cos\phi & -\sin\phi & 0\\
                    \sin\phi & \cos\phi & 0\\
                    0 & 0 & 1
                    \end{pmatrix}, \quad \phi(D_l, l, b).
\end{split}
\end{equation}
Here $\mathbf{v}^\mathbb{R}$ is the linear velocity relative to the Sun in the Cartesian coordinate, $\mathbf{v}_\odot^\mathbb{R}$ is the Sun's velocity, and $\phi$ is the angle in the cylindrical coordinate. Then, we can rotate from the Cartesian coordinate to galactic coordinate $(r, l, b)$
\begin{equation}
\begin{split}
    P_\text{BH}(\mathbf{v}^G|D_l, l, b) &= \frac{1}{(2\pi)^{3/2}\det\pmb{\Sigma}}\exp\left(-\frac{1}{2}{\mathbf{v}^G}^T\mathbf{R}_2\pmb{\Sigma}^{-2}\mathbf{R}_2^T\mathbf{v}^G\right)\\
    \mathbf{R}_2 &= \begin{pmatrix}
                    \cos b\cos l & \cos b\sin l & \sin b\\
                    -\sin l & \cos l & 0\\
                    -\sin b\cos l & -\sin b\sin l & \cos b
                    \end{pmatrix}\mathbf{R}_1.
\end{split}
\end{equation}
Here $\mathbf{v}^G$ is the linear velocity in galactic coordinate. One more rotation brings the velocity into equatorial coordinate $(r, \alpha, \delta)$
\begin{equation}
\begin{split}
    P_\text{BH}(\mathbf{v}^E|D_l, l, b) &= \frac{1}{(2\pi)^{3/2}\det\pmb{\Sigma}}\exp\left(-\frac{1}{2}{\mathbf{v}^E}^T\mathbf{R}_3\pmb{\Sigma}^{-2}\mathbf{R}_3^T\mathbf{v}^E\right),\\
    \mathbf{R}_3 &= \begin{pmatrix}
                    1 & 0 & 0\\
                    0 &\cos\psi & -\sin\psi\\
                    0 & \sin\psi & \cos\psi
                    \end{pmatrix}\mathbf{R}_2,\quad \psi(l, b).
\end{split}
\end{equation}
Here $\mathbf{v}^E$ is the linear velocity in equatorial coordinate. Finally, we can integrate out the radial velocity to obtain the PDF of the velocity in the perpendicular component $\mathbf{v} = (v_{\alpha}, v_\delta)^T$
\begin{equation}
\begin{split}
    P_\text{BH}(\mathbf{v}|D_l, l, b) &= \int_{-\infty}^\infty \mathrm{d}v_r
    P_\text{BH}(\mathbf{v}^E|D_l, l, b)\\
    &= \frac{1}{2\pi\det\pmb{\Sigma}\sqrt{a_{11}}}\exp\left(-\frac{1}{2}\mathbf{v}^T\mathbf{A}\mathbf{v}\right),\\
    \mathbf{R}_3\pmb{\Sigma}^{-2}\mathbf{R}_3^T &= \begin{pmatrix}
                    a_{11} & a_{12} & a_{13}\\
                    a_{12} & a_{22} & a_{23}\\
                    a_{13} & a_{23} & a_{33}
                    \end{pmatrix},\\
    \mathbf{A} &= \begin{pmatrix}
                 a_{22} - \frac{a_{12}^2}{a_{11}} & a_{23} - \frac{a_{12}a_{13}}{a_{11}}\\
                 a_{23} - \frac{a_{12}a_{13}}{a_{11}} & a_{33} - \frac{a_{13}^2}{a_{11}}
    \end{pmatrix}.
\end{split}
\end{equation}
Finally, we can perform a change of variable from $\mathbf{v}$ to obtain the conditional PDF $P(\pmb{\mu}_l|D_l, l, b)$
\begin{equation}
    P_\text{BH}(\pmb{\mu}_l|D_l, l, b) = \frac{D_l^2}{2\pi\det\pmb{\Sigma}\sqrt{a_{11}}}\exp\left(-\frac{D_l^2}{2}\pmb{\mu}_l^T\mathbf{A}\pmb{\mu}_l\right).
\end{equation}
A sample of this conditional PDF at $(l, b) = (270^\circ, 0^\circ)$ and $D_l = 1$ kpc is shown in figure~\ref{fig:pm_dist_pdf}\\

\section{Photometric lensing signal of astrophysical BHs in \textit{Gaia}}
\label{sec:photometric}

\begin{figure}
    \centering
    \includegraphics[width = \textwidth]{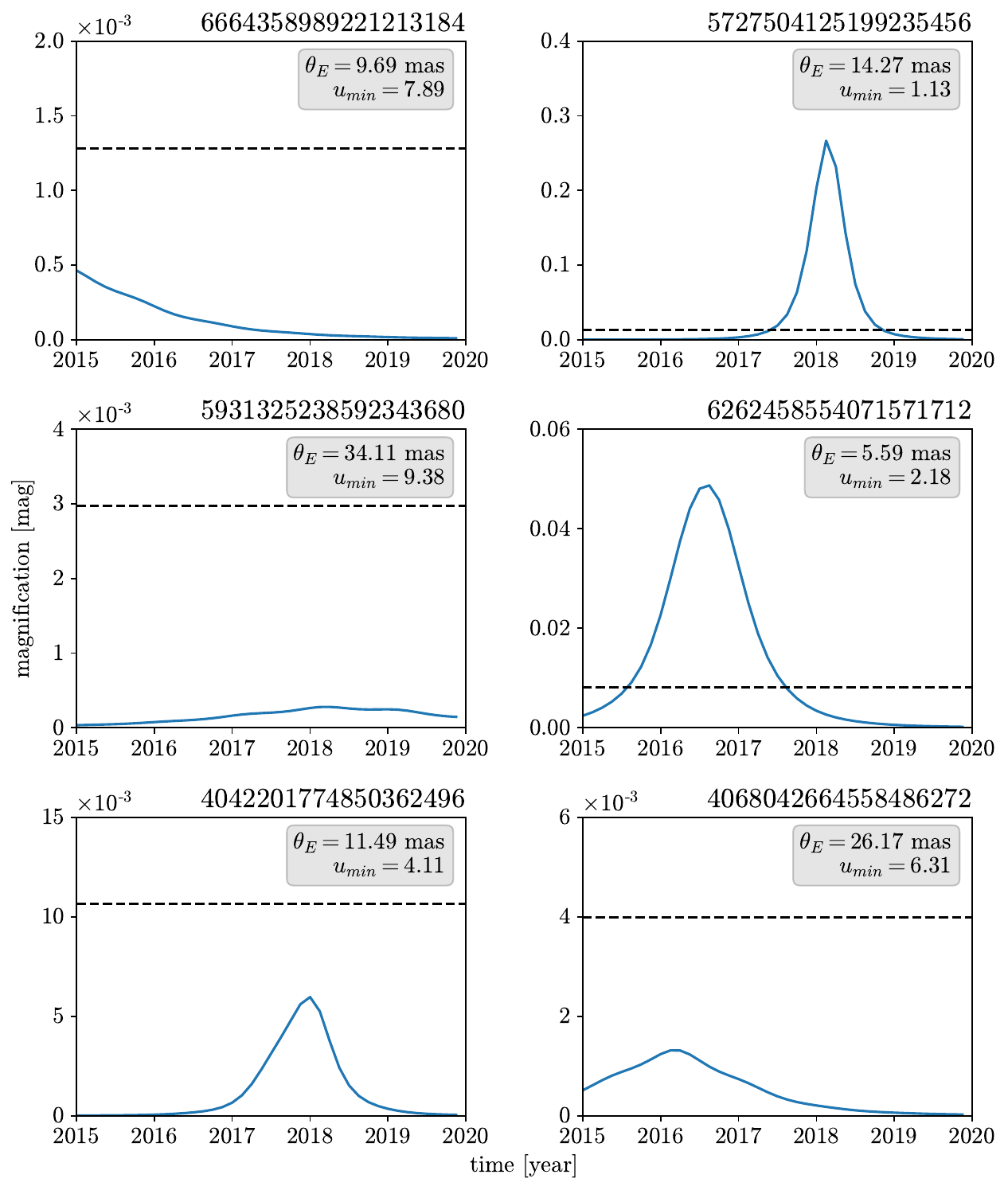}
    \caption{Light curves from the 6 lensing events described in section~\ref{sec:bh_catalog}. The horizontal black solid lines are the \textit{Gaia} photometric uncertainty of each source. Note that only two sources (5727504125199235456 and 6262458554071571712) have magnification larger than the \textit{Gaia} photometric sensitivity. The Einstein radius and the minimum dimensionless impact parameter of each event are shown in the legend. \nblink{PaperPlots/mag.ipynb}}
    \label{fig:photometric}
\end{figure}

We can calculate the total magnification of a point-like background star due to a point-like foreground lens by summing up the magnification of the two images in eq.~\eqref{eq:two_soln_lensing_mag} when the two lensed image are not resolved independently using \eqref{eq:total_mag}.
Using this equation, we calculate the light curves of the 6 candidate lensing events found in our mock catalog (see section~\ref{sec:bh_catalog}). These light curves are shown in figure~\ref{fig:photometric}. The horizontal black dashed line is the photometric uncertainty per transit for each of the background stars taken from ref.~\cite{prusti2016gaia}. The Einstein radius and the minimum dimensionless impact parameter for each event in shown in the plot as well. We can see that the maximum brightening is below the \textit{Gaia} sensitivity for all but 2 (5727504125199235456, 6262458554071571712) sources. And for the 4 sources that pass the $3\sigma$ acceleration fit, only one (5727504125199235456) has magnification larger than the \textit{Gaia} photometric uncertainty.

The \textit{Gaia} Photometric Alert System \cite{Hodgkin2021} will likely discover lensing events similar to 5727504125199235456. In fact, there is already a successful detection of a lensing event in Gaia with ID Gaia16aye \cite{Wyrzykowski_2020}. This event was flagged using the \textit{Gaia} Photometric Alert System and later confirmed to be a foreground lens consisting of a binary star system via \textit{Gaia} astrometry coupled with ground-based photometry follow-up. However, this work demonstrates that many significant lensing events will evade a photometric alert system and only be detectable via astrometry.

\section{Derivation of analytic constraint projection}
\label{sec:pbh_estimate}
Suppose that stars in the \textit{Gaia} catalog are distributed evenly and are stationary at infinity. A lens with velocity $v$ will sweep through an area of $2v\tau b_\text{min}$. Thus, the expected minimum impact parameter of all lens is
\begin{equation}
    \langle b_\text{min}\rangle = \frac{3M_l}{2v\tau N_*\rho_\text{DM}D_lf_l},
\end{equation}
where $N_*$ is the number of stars in the \textit{Gaia} catalog. For this event to be a blip we require that $\langle b_\text{min}\rangle < v\tau$. Plugging in $N_* = 1.4\times 10^9$, $\rho_\text{DM} = 10^{-2}~M_\odot\,\text{pc}^{-3}$, $D_l = 10$~kpc, we arrive at the rightmost branch of the analytic estimate:
\begin{equation}
 f_l\geq \frac{3M_l}{2(v\tau)^2 N_*\rho_\text{DM}D_l}.
\end{equation}

On the other end, the $\Delta\chi^2$ used for the event selection is a proxy of SNR$^2$, which can be parameterized by
\begin{equation}
    \text{SNR}^2=\frac{\delta_\text{max}^2}{\sigma_\theta^2}\frac{N_\text{obs}b_\text{min}}{v\tau} = \left(\frac{4GM_l}{c^2\sigma_\theta}\right)^2\frac{N_\text{obs}}{b_\text{min} v\tau}.
    \label{eq:SNR}
\end{equation}
For the SNR to reach some threshold, we then arrive at the expression:
\begin{equation}
    f_l\geq\left(\frac{\text{SNR}\,c^2\sigma_\theta}{4G}\right)^2\frac{3}{2N_*\rho_\text{DM}D_lN_\text{obs}M_l}.
    \label{eq:kvt_SNR}
\end{equation}
Accounting for the look-elsewhere effect and the average lens distance for significant events, we use $\text{SNR} = 15$ and $D_l = 1$ kpc, which yields the left branch of the red dashed-dotted analytic estimate in figure~\ref{fig:PBH}, closer to the simulation done in this work.

For strong lensing that saturates the astrometric deflection, eq.~\eqref{eq:SNR} is modified as
\begin{equation}
    \text{SNR}^2=\frac{\theta_E^2}{8\sigma_\theta^2}\frac{N_\text{obs}D_l\theta_E}{v\tau} = \left(\frac{4GM_l}{c^2D_l}\right)^{3/2}\frac{N_\text{obs}D_l}{8\sigma_\theta^2v\tau},
    \label{eq:strong}
\end{equation}
The expected distance to the closest lens $\langle D_l\rangle$ can be expressed as
\begin{equation}
    \langle D_l\rangle = \left(\frac{3M_l}{4\pi\rho_\text{DM}f_l}\right)^{1/3}.
\end{equation}
Plug this back into eq.~\eqref{eq:strong} to get the sharp cutoff in the left branch of the red-solid curve in figure~\ref{fig:PBH}.

\begin{figure}
    \centering
    \includegraphics[width = .7\textwidth]{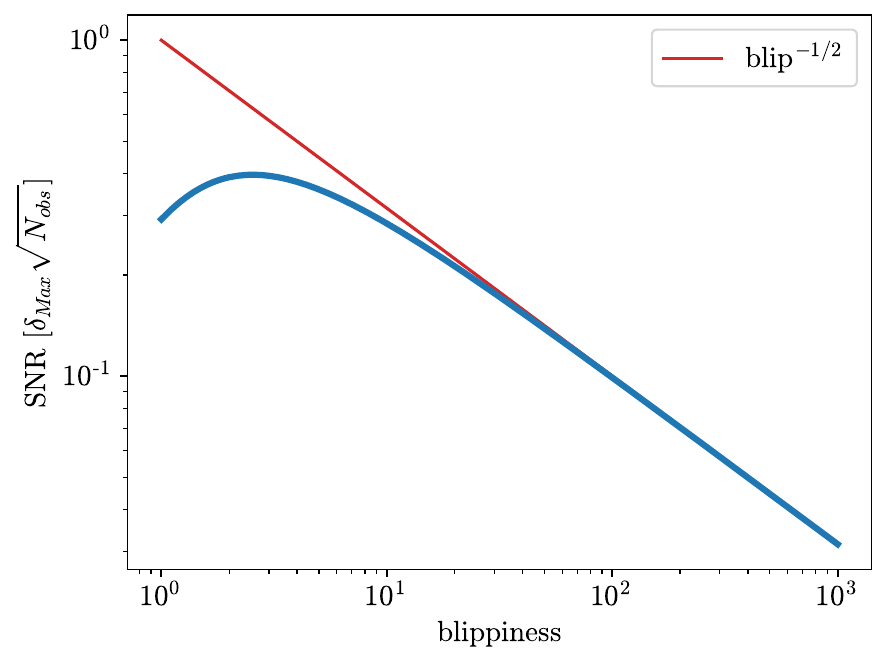}
    \caption{The relationship between SNR to blippiness. The red curve shows the relation adopted in ref.~\cite{van2018halometry} as shown in the left branch of the red dashed-dotted curve in figure~\ref{fig:PBH} and in eq.~\eqref{eq:kvt_SNR}, which uses the maximum deflection $\delta_{\text{max}}$ for calculating SNR. The blue curve shows the relation using the difference of maximum deflection and minimum deflection throughout the mission time $\delta_\text{max} - \delta_\text{min}$ for calculating SNR, as is the relation used in the horizontal branch of the red solid curve in figure~\ref{fig:PBH} and in eq.~\eqref{eq:flat}. Here we can see that the scaling changes for blippiness $\lesssim10$ and the SNR remains approximately constant in thie regime. \nblink{PaperPlots/blip.ipynb}}
    \label{fig:slow}
\end{figure}

Another thing we discovered is that eq.~\eqref{eq:SNR} only applies when the blippiness is large $(\gtrsim 10)$ because of the definition of $\delta_\text{max}$, which should be $\delta_\text{max} - \delta_\text{min}$ for calculating $\Delta\chi^2$. For events with large blippiness, $\delta_\text{min} \approx 0$ so eq.~\eqref{eq:SNR} is valid. However, as figure~\ref{fig:slow} shows, events with small blippiness $(\lesssim10)$ do not follow this relation, becoming almost independent of blippiness, which we parametrize as the following:
\begin{equation}
    \frac{\text{SNR}^2}{0.4^2}=\frac{\delta_\text{max}^2}{\sigma_\theta^2}N_\text{obs} = \left(\frac{4GM_l}{c^2 b_\text{min}\sigma_\theta}\right)^2N_\text{obs},
    \label{eq:flat}
\end{equation}
where $0.4$ is the peak of the blue curve in figure~\ref{fig:slow}. This gives the constraint:
\begin{equation}
    f_l\geq\frac{3c^2\sigma_\theta}{8Gv\tau N_*\rho_\text{DM}D_l}\frac{\text{SNR}}{0.4\sqrt{N_\text{obs}}},
\end{equation}
which is the horizontal branch of the red-solid curve in figure~\ref{fig:PBH}.

\end{document}